\documentclass[
	twocolumn,
	amsmath,
	amssymb,
	prx,
	aps,
	nofootinbib,showpacs,superscriptaddress
	]{revtex4-2} 
\usepackage{graphicx}
\usepackage{xcolor}
\usepackage{amsmath}
\usepackage{amssymb}
\usepackage{mathtools}
\usepackage{booktabs}
\usepackage{titlesec}
\usepackage{float}

\begin{document}

\title{Path-dependency and emergent computing under vectorial driving}

\author{C.M. Meulblok}
\affiliation{Huygens-Kamerlingh Onnes Lab, Universiteit Leiden, P.O.~Box~9504, NL-2300 RA Leiden, Netherlands}
\affiliation{AMOLF, Science Park 104, 1098 XG Amsterdam, Netherlands}

\author{A. Singh}
\affiliation{Huygens-Kamerlingh Onnes Lab, Universiteit Leiden, P.O.~Box~9504, NL-2300 RA Leiden, Netherlands}
\affiliation{AMOLF, Science Park 104, 1098 XG Amsterdam, Netherlands}

\author{M. Labousse}
\affiliation{Huygens-Kamerlingh Onnes Lab, Universiteit Leiden, P.O.~Box~9504, NL-2300 RA Leiden, Netherlands}
\affiliation{AMOLF, Science Park 104, 1098 XG Amsterdam, Netherlands}
\affiliation{Gulliver, CNRS, ESPCI Paris, Universit\'{e} PSL, 75005 Paris, France}

\author{M. van Hecke}
\affiliation{Huygens-Kamerlingh Onnes Lab, Universiteit Leiden, P.O.~Box~9504, NL-2300 RA Leiden, Netherlands}
\affiliation{AMOLF, Science Park 104, 1098 XG Amsterdam, Netherlands}

\date{\today}

\begin{abstract}
The sequential response of frustrated materials—ranging from crumpled sheets and amorphous media to metamaterials—reveals their memory effects and emergent computational potential. Despite their spatial extension, most studies rely on a single global stimulus, such as compression, effectively reducing the problem to scalar driving.
Here, we introduce vectorial driving {of frustrated materials} by applying multiple spatially localized stimuli to explore path-dependent, sequential responses. We uncover a wealth of phenomena absent in scalar driving, including non-Abelian responses, mixed-mode behavior, and chiral loop transients. {We show that fold singularities connect three states—ancestor, descendant, and sibling. This 
recurring pattern serves as the elementary building block of all sequential paths.} 
{We then introduce three levels of description of sequential, path-dependent responses.
At the most fundamental level, {path-dependent transition graphs (pt-graphs)} and strain maps capture the response under arbitrary vectorial driving and connect pathways to the underlying singularities. They provide a complete description analogous to {transition graphs (t-graphs)} for scalar driving.
However, as pt-graphs and strain maps become unwieldy for high-dimensional driving, we introduce b-graphs{--graphs whose nodes and transitions encode the systems response to binarized vectorial driving. These} present a less complete but much simpler second-level description by 
{restricting attention to binary input and their induced transitions}. 
{The resulting} graphs form subgraphs of the pt-graph and naturally connect to sequential computing and 
Boolean circuits. 
We show that a single sample can encode multiple b-graphs that are selectable through the driving strength and can be reprogrammed via additional inputs.
Finally, we introduce graph-based motifs that enable a systematic analysis of b-graphs and provide a practical tool for characterizing the statistical response of very high-dimensional systems, where a full pt- or b-graph description is neither feasible nor particularly insightful. As statistical measures of pathway complexity, these motifs can be obtained in systems of any size or complexity.}
Our work paves the way for strategies to explore, harness, and understand complex materials and memory, while advancing embodied intelligence and {\em in-materia} computing.
\end{abstract}

\maketitle

\section{INTRODUCTION}

A central goal of condensed matter physics is to describe and understand how materials respond to external stimuli. This is particularly challenging for frustrated, multistable materials, such as amorphous solids, crumpled sheets, or frustrated metamaterials \cite{SlotterbackPRE2012,PaulsenPRL2014,AdhikariEPJ2018,KeimPRR2020,RegevPRE2021,MatanPRL2002,AharoniNatMat2010,LahiniPRL2017,BensePNAS2021,ShohatPNAS2022,MeeussenNat2023,FlorijnPRL2014,GuoNat2023}. Their state does not only depend on the current driving, but also on the driving history,
leading to a wealth of memory effects which classify materials and give insight into their underlying physics \cite{KeimRMP2019,PaulsenAR2025}.
Examples of such memory effects include
measuring after how many driving cycles the response becomes periodic
\cite{PaulsenPRL2014,BensePNAS2021,LindemanSciAdv2021,RegevPRE2013,SchreckPRE2013,RoyerPNAS2015,KawasakiPRE2016,LavrentovichPRE2017,NagasawaSM2019,YehPRL2020,KwakernaakPRL2023}, or whether the material returns to a previous state when the driving returns to an earlier extremum \cite{BensePNAS2021,ShohatPNAS2022,KeimSciAdv2021,LiuPNAS2024,MunganAHP2019}. More complex memory effects can be leveraged to realize {\em in materia} computing, where the materials state represents the outcome of a calculation, and the initial state and driving history {are} its input \cite{KwakernaakPRL2023,LiuPNAS2024}. 

So far, studies of memory and computing have focused on driving with a single driving field, such as global compression or shear. 
For such {\em scalar}, quasistatic driving of athermal materials, 
transition graphs are emerging as a powerful tool for characterizing memory effects \cite{RegevPRE2013,BensePNAS2021,LindemanSciAdv2021,LiuPNAS2024,KeimSciAdv2021,MunganAHP2019,MunganPRL2019,PaulsenPRSA2019,TerziPRE2020,HeckePRE2021,MelanconAdvFuncMat2022,JulesPRR2022,DingJCP2022,LindemanSciAdv2025}. These represent states as nodes and irreversible transitions under sweeps of the driving parameter $\varepsilon$ as edges. They encode the response to any sequential driving protocol, thus allowing {precise characterization of} memory effects and emergent computational capabilities \cite{PaulsenPRSA2019,LiuPNAS2024}.

However, the spatial extension of complex materials naturally 
leads to the question of spatially textured driving. For example, one can push a crumpled sheet or metamaterial at various locations in various orders
\cite{GuoNat2023,FlorijnPRL2014}. We refer to this as {\em vectorial} sequential driving, with each component of $\vec{\varepsilon}$ corresponding to, e.g., the compression or strain at a given location. For strongly 
nonlinear responses associated with complex materials, superposition no longer describes the response. Instead 
the driving {\em path} from 
$\vec{\varepsilon}_1$ to $\vec{\varepsilon}_2$ 
plays a crucial role, 
leading to sequential responses, memory effects, and computational capabilities that have no counterpart in scalar driving. 
We note that sequential vectorial driving itself is not new - think of the Carnot cycle - 
and path-dependencies 
have been studied extensively for materials that exhibit
plasticity \cite{MozaffarPNAS2019}. Here we focus on 
{developing a coherent and practical framework to characterize the} path-dependencies of multistable, athermal materials that are driven quasistatically.

We address the following broad questions:
What new phenomena, memory effects, and computational capabilities can be
accessed with vectorial driving? 
How can path-dependent sequential responses be described?
How to deal with the complexity associated with 
high-dimensional driving?

\begin{figure*}[t]
	\centering
	\includegraphics[]{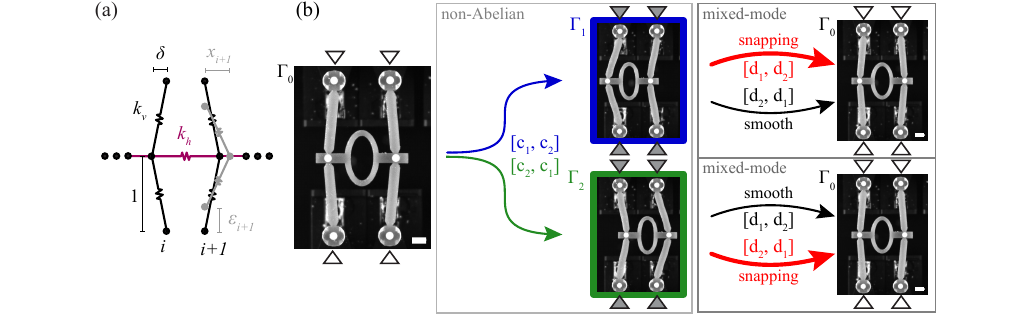}
	\caption{ Geometry and phenomenology. 
  (a) Metamaterial consisting of a staggered chain of $n$ left-right symmetry broken units (here $i$ and $i+1$). Each unit consists of two nearly vertical elastic elements (spring constant $k_v$), and neighboring units are coupled with elastic elements (spring constant $k_h$). The material is driven by a vectorial strain $\vec{\varepsilon}$, consisting of vertical strains $\varepsilon_i$
  applied independently by compressing each unit.
  We non-dimensionalize lengths by the distance between the endpoints of each unit at rest, {$2\ell_0 =2$}, and specify the configuration by the (non-dimensional) horizontal displacements $\vec{x}=(x_1,x_2,\dots)$.
  The essential design parameters are the dimensionless symmetry breaking $\delta=x_i(\varepsilon_i=0)$ and spring ratio $k:=k_h/k_v$. 
  (b) Left: top view of the experimental sample A in its rest configuration $\Gamma_0$ ($n=2$, $k\approx0.09$ and $\delta\approx0.08$, scale bar: 5~mm; see appendix~A for details). The central ellipse materializes a soft horizontal spring, and the top and bottom of the vertical beams connect to the strain apparatus, where white and gray triangles indicate relaxed and compressed boundaries. Middle and Right: illustration of two hallmarks of path dependency, under sequential compression, respectively decompression of the first and second unit ($c_1,c_2$ resp. $d_1,d_2$). Middle: a non-Abelian response where the final configuration
  $\Gamma_1$ obtained by compression path $\left[c_1 ,c_2\right]$ (green) differs from
  configuration
  $\Gamma_2$ obtained by compression path $\left[c_2 ,c_1\right]$ (blue). Right: Mixed mode response where the response is smooth (black) or non-smooth (red) depending on the sequence of the decompression path. {See Movie~1 for the real-space evolution of these path dependent responses.}
	}\label{fig:PathDependency}
\end{figure*}

To address these questions, we combine experiments with simulations and theory to probe the path-dependent sequential response of metamaterials based on chains of coupled nonlinear units (Fig.~\ref{fig:PathDependency}a). 
These chains naturally allow vectorial driving by compressing the individual units. By designing the units as oppositely biased elastic beams, the metamaterial forms a 
chain of tunable double-well elements with antagonistic nearest-neighbor interactions, which
captures generic features of the path-dependent response of complex materials. 

We start by investigating a short chain comprising two coupled units (Sec.~\ref{sec:system}). By comparing the evolution of a given initial state
under different driving paths with
the same initial and final driving, we uncover two hallmarks of path-dependency: a 
{\em non-Abelian} response, where the final state depends {on} the path, and a {\em mixed-mode} {response}, where the nature of the evolution - {reversible, continuous or irreversible, discontinuous}
- depends on the path. This simple example also illustrates that two-dimensional driving already {has} behaviors unique to vectorial driving.
We show that fold singularities govern
path dependency, and that paths crossing them link three states - ancestor, descendant, and sibling - leading to ADS-triplets that form the elementary unit of path-dependency. 

Based on these triplets, we present a general framework 
which allows to encode the sequential response to {\em any} vectorial driving protocol (Sec.~\ref{sec:General}). This framework uses strain maps to label the location of fold singularities in driving space, and path-transition {graphs} (pt-graphs), which extend the notion of t-graphs developed for scalar driving. 
We show that while pt-graphs under scalar driving can be mapped to t-graphs, pt-graphs for {higher-dimensional} driving
have qualitatively distinct features.

To enable the experimental characterization of path dependencies under high-dimensional driving (i.e., {many} inputs), we introduce binarized on/off driving protocols. These protocols focus on paths connecting strains $\vec{\varepsilon} := \varepsilon^M \vec{b}$, where $\varepsilon^M$ is the overall strain and 
$\vec{b}$ is a binary vector (Sec.~\ref{sec:binary}).
The response can then be captured by `b-graphs' that are simpler than the pt-graphs, and facilitate 
a natural link {to} sequential computing {and}
Boolean circuits.
Strikingly, we show that a single sample embeds multiple circuits which can be selected by the strain scale $\varepsilon^M$. {
We then analyze larger samples and determine the b-graphs as a function of the driving scale $\varepsilon^M$ for a four-unit chain (Sec.~\ref{sec:fourunit})}.

To address the combinatorial complexity of b-graphs under high-dimensional vectorial driving, we introduce motifs. These can be seen as
low-dimensional building blocks of the b-graph, but also can be determined independently of the full b-graph, even in systems of arbitrary size.
We first discuss 
edge motifs, which consider driving paths restricted to one-dimensional subspaces of $\vec{\varepsilon}$, and we show that these are composed of elementary ADS-patterns. 
Loop motifs, associated with loop-like paths in two-dimensional subspaces, probe essential aspects of the sequential response to vectorial driving and exhibit substantial diversity—our four-unit example shows 13 distinct types.
They capture a broad range of path dependencies, 
including the elementary mixed-mode and non-Abelian responses. 

We analyze the loop motifs
to uncover emergent memory effects and computational capabilities
of our samples (Sec.~\ref{sec:motifs}). We {observe} chiral loop transients where the transient time - i.e., the number of driving loops before the response becomes periodic - depends on the orientation of the loop. This exemplifies the notion of memory effects that have no counterpart for scaling driving.
Finally, by splitting the binarized driving at the four inputs of our sample {into} two for computation, and two for 'programming', we show that a single sample can be reprogrammed to mimic a variety of sequential Boolean circuits - on top of the variety that is accessible by varying the magnitude of these drivings.

{Finally, to illustrate that our framework is applicable to systems with disordered pathways, we study compressed corrugated sheets
(Sec.~\ref{sec:disorder}). Despite the absence of a simple parametrization of the
states, and only using two-dimensional driving, we show that being able to determine the location of irreversible transitions and compare different configurations is sufficient to describe the pathways in our framework. In particular, these systems exhibit complex motifs, including edge and loop transient structures not seen in the metamaterial chain. While obtaining the full pt-graph or set of b-graphs is challenging for such systems, this demonstrates that motifs are a powerful tool
to systematically uncover and characterize the complex
path dependencies under vectorial driving of multistable
systems.}

Together, our findings uncover the elementary motifs that underlie path-dependent behavior in multi-stable systems.
While we focus on mechanics, our approach can be applied to any multi-stable system under athermal and quasi-static driving.
Vectorial driving thus opens new avenues to characterize frustrated systems, to observe new classes of 
multi-dimensional responses and memories, and to develop strategies for {powerful} {\em in-materia} computing.

\section{SYSTEM AND PHENOMENOLOGY} \label{sec:system}

We introduce chain-like multistable metamaterials that readily enable compression at multiple locations,
along with a corresponding minimal spring model (Sec.~\ref{sec:methods}).
We investigate a short chain and compare the 
response to pairs of strain paths that have the same initial and final strains (Sec.~\ref{sec:PDs}). We identify two archetypical scenarios for the path-dependent response under vectorial sequential driving:
First, a {\em non-Abelian} response, where the final configuration depends on the strain path (Sec.~\ref{sec:nonabelian}); second,
a {\em mixed-mode} response, where the response is either smooth or non-smooth, depending on the strain path (Sec.~\ref{sec:mm}). Both can be understood to arrive from a fold bifurcation, which links three configurations that we term
ancestor, descendant and sibling (Sec.~\ref{sec:sn}).

\subsection{Metamaterial, Strain Protocols and Model} \label{sec:methods}

{\em Design.---} We aim to realize a metamaterial that naturally allows compression at multiple locations, is scalable, can be multistable, and features path-dependent responses. 
To meet these requirements, we investigate a chain-like geometry of $n$ units that each have a strongly nonlinear response and can become bistable. We design such a geometry based on a spring chain, consisting of $n$ vertical units that each contain two equal springs meeting in a central node that acts as a hinge.
Neighboring units are coupled by a horizontal spring between these nodes. The horizontal positions of the ends of the units are
fixed such that in their relaxed state, the units are not perfectly vertical but alternately kink left and right (Fig.~\ref{fig:PathDependency}a).
This configuration ensures a deterministic response when
the ends of the units are compressed vertically.
Moreover, if one unit is compressed sufficiently to flip the orientation of the kink of its neighbor, we anticipate {that} first compressing unit $i$ and then $i+1$ results in a different configuration {than} first compressing unit $i+1$ and then $i$: this multistable metamaterial has a 
path dependent response (Fig.~\ref{fig:PathDependency}b).

{\em Sample Geometry.---} Experimentally, we realize metamaterials based on this design
in flexible silicone rubber using standard 3D printing and molding techniques {(see appendix~A)}. Each unit $i$ consists of two connected, nearly vertical elastic beams of stiffness $k_v$ which are coupled by horizontal beams of spring constant $k_h$; the ratio
$k:= k_h/k_v$ is an important parameter.
The beams have tapered connections to minimize torques. As we are interested in the limit where $k \ll 1$, we realize the soft horizontal springs using an elliptical annulus (Fig.~\ref{fig:PathDependency}b).
The distance between the endpoints of each unit at rest, {$2\ell_0$,
defines the length scale $\ell_0$ that we use to non-dimensionalize all lengths.}
The units are pre-kinked over a (non-dimensional) distance $\delta$,
such that their rest positions
(defined as the horizontal deviation from the unit's midpoint) are
$x_i=(-1)^{i+1}\delta$ (this requires the beams rest length to be
$1 + \delta^2/2$).
We focus here on geometries with $\delta\approx0.08$ and $k\approx0.09$, and create samples with $n=2$ and $n=4$ units (see appendix~A).

{\em Strain protocols.---} We use a custom-made device to control the spatially localized compressive strains $\varepsilon_i$ {with accuracy $\sim 10^{-3}$ by symmetrically shortening the end-to-end distances $2 \ell=2\ell_0(1-\varepsilon_i)$ such that all nodes remain horizontally aligned. }
We use quasi-static driving (strain rate is $5\cdot 10^{-4}$ s$^{-1}$) 
and track the real-space configuration of the structure given by
$\vec{x}=(x_1,x_2,\dots)$
using a CCD camera, resulting in an accuracy in $x_i$ of $0.002$
(see appendix~A). Together, our {setup} allows for automated driving protocols and precise measurements of the configurational response.

We consider strain paths $P$ that quasistatically ramp the input strains $\vec{\varepsilon}=(\varepsilon_1,\varepsilon_2,\dots)$ between sequences of specific points in strain-space.
While we also consider general driving protocols, we often employ 'binary' paths that consist of sequences of elementary compressions
$c_i$ (increase of $\varepsilon_i$
from {a minimum value $\varepsilon_{i}^{m}$ to a maximum value $\varepsilon_{i}^{M}$)}
and decompressions $d_i$ (decrease of $\varepsilon_i$ from $\varepsilon_{i}^{M}$ to $\varepsilon_{i}^{m}$)
while the other strains are kept fixed.
In many cases we take $\varepsilon_{i}^{m} = 0$ and $\varepsilon_{i}^{M}=\varepsilon^M$, so that $\varepsilon^M$ sets an overall strain scale. Finally, we
denote sequences of compression and decompression steps using square brackets, e.g., $[c_1,c_2,d_1,d_2]$ represents a loop-like compression path (read from left to right).

{\em Minimal Model.---} In addition to the experiments, we numerically explore a spring model. We model each beam as a linear spring with freely rotating hinges, use equidistance vertical compression points, and break the left-right symmetry by taking horizontal springs with alternating rest length $1 + (-1)^i2\delta$ (Fig.~\ref{fig:PathDependency}a).
Expanding the elastic energy to the lowest order in $x_i$ yields:
\begin{equation}
	E = \sum_{i=1}^{n} \left( \frac{x_i^2}{2} - \gamma_i \right)^2 + \sum_{i=1}^{n-1} \frac{k}{2} \left( x_i - x_{i+1} - (-1)^i2\delta \right)^2 ~,
	\label{eq:energy}
\end{equation}
where $\gamma_i := \varepsilon_i + \delta^2/2 - \varepsilon_i^2/2 \approx \varepsilon_i + \delta^2/2$.
The first term describes the tuneable quartic potential of each unit, and the second term describes the interactions and lifts the units degeneracy to make the response deterministic.
In the limit of small $\delta$, one can rescale the parameters
to eliminate $\delta$,
leaving the effective interaction strength $k$ as the {main system-dependent} parameter (see appendix~B).

\subsection{Path dependency} \label{sec:PDs}

Here, we explore path dependency for a
two-unit sample (A), and demonstrate two paradigmatic forms of path-dependency which have no counterpart 
under scalar driving: {a non-Abelian} response and {a mixed-mode} response.

\subsubsection{Non-Abelian Response}\label{sec:nonabelian}

We consider a pair of sequential driving paths $P_1: [ c_1,c_2 ] $ and $P_2: [ c_2,c_1 ]$, with $\varepsilon^m_i=0$ and $\varepsilon^M_i=\varepsilon^M=0.03$ (Fig.~\ref{fig:PathDependency}b, Movie~1).
The sample is initially relaxed $(\vec{\varepsilon} = 0)$ in the unique rest configuration $\Gamma_0$, such that the driving paths $P_1$ and $P_2$
produce smooth deformations until the final configurations $\Gamma_1=P_1(\Gamma_0)$ and $\Gamma_2=P_2(\Gamma_0)$ are reached.
We compare the final configurations \( \Gamma_1 \) and \( \Gamma_2 \) and find \( \Gamma_1 \neq \Gamma_2 \), demonstrating path dependence (Fig.~\ref{fig:PathDependency}b, Movie~1).
Since \( \Gamma_1 \) and \( \Gamma_2 \) experience the same strain \( \vec{\varepsilon} = (\varepsilon^M, \varepsilon^M) \), this non-Abelian response necessitates bistability. Similar non-Abelian behaviors have recently been observed in various other strongly nonlinear, multistable metamaterials under sequential compression \cite{GuoNat2023,SiroteNatComm2024,FlorijnPRL2014}.

In our system, which was designed with such
behavior in mind, this non-Abelian response is observed 
for a large range of strains (see Sec.~\ref{sec:binary}).
This behavior can intuitively be understood
from the left ($x_i<0$) or right ($x_i>0$) leaning orientations of the units, although we track the position of all coordinates $x_i$. 
Consider the evolution of the real-space configuration \( \{x_i\} \) during \( P_1 \) (Fig.~\ref{fig:PathDependency}b; see Movie S1). In the first stage ($c_1$), unit 1 compresses leftward, pulling unit 2 to lean left as well. Further compression of unit 2 in stage two ($c_2$) pushes both units left, making the resulting configuration \( \Gamma_1 \) left-leaning. By contrast, in \( P_2 \), compression of unit 2 shifts unit 1 rightward, yielding a right-leaning \( \Gamma_2 \) (Fig.~\ref{fig:PathDependency}b).
We note, however, that such orientation changes are not required to observe a non-Abelian response: intermediate values of \( \varepsilon^M \approx 0.011 \) 
lead to a more subtle non-Abelian regime without changes in the {unit} orientations (Appendix~C).

\subsubsection{Mixed-mode response} \label{sec:mm}
Next, we follow the configurations during the decompression paths $P_3:[d_1, d_2]$ and $P_4 : [d_2, d_1]$ (Fig.~\ref{fig:PathDependency}b, Movie~1). 
Starting from either of the configurations in the bistable regime
at $\vec{\varepsilon}=(\varepsilon^M,\varepsilon^M)$, both paths
$P_3$ and $P_4$ lead to the same configuration, i.e., $P_3(\Gamma_1)=P_4(\Gamma_1)=\Gamma_0$ and $P_3(\Gamma_2)=P_4(\Gamma_2)=\Gamma_0$. 
However, whereas $P_4(\Gamma_1)$ leads to a smooth deformation,
$P_3(\Gamma_1)$ features a non-smooth, irreversible transition --- and similarly for $P_3(\Gamma_2)$ and $P_4(\Gamma_2)$
(Fig.~\ref{fig:PathDependency}b, Movie 1)\footnote{While here the discontinuous transition is associated with decreasing strain, we will later also encounter discontinuous transitions under increasing strain.}. 
This mixed-mode response represents a different type of
path-dependency than the non-Abelian response.

\subsubsection{Fold bifurcation}\label{sec:sn}

\begin{figure}[bt]
	\centering
	\includegraphics[]{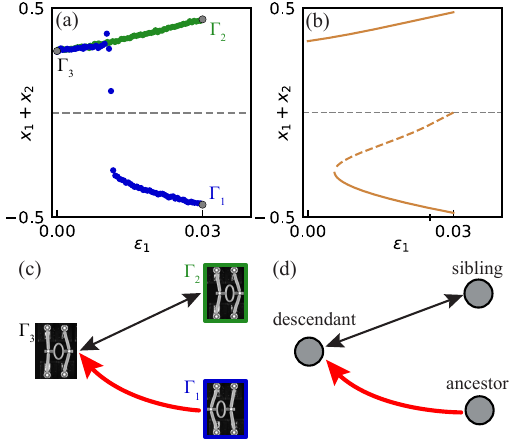}
	\caption{ Fold bifurcation and ADS-triplet
  (a-b) Evolution of the configuration $\vec{x}$ of sample A as a function of strain $\varepsilon_1$ which is lowered from 0.03 to 0, 
  projected on the collective coordinate $x_1+x_2$; here we fix $\varepsilon_2=\varepsilon^M=0.03$. (a) Experimental data for sample A; we initialize the sample in configurations $\Gamma_1$ and $\Gamma_2$ using the strain paths shown in Fig.~1b.
  {The error bars of the measured points are smaller than the symbol size.} 
  (b) Simulations of the corresponding spring model, where $\varepsilon_1$ 
  is both increased and decreased, showing stable (full) and unstable (dashed) solutions, thus revealing a fold singularity. (c-d) Graph representation of the general structure underlying path dependency: three configurations (ancestor, descendant and sibling) are connected by reversible (black) and irreversible (red) transitions; $\Gamma_1$ is the ancestor, $\Gamma_3$ the descendant, and $\Gamma_2$ the sibling. {See Movie~2 for the real-space evolution.}
	}\label{fig:fold}
\end{figure}

To understand the root cause of the non-Abelian and mixed mode response, we follow the evolution of the configurations $\Gamma_1$ and $\Gamma_2$ between strains 
{starting from $(\varepsilon^M,\varepsilon^M)$ to $(0,\varepsilon^M)$}
(Fig.~\ref{fig:fold}a, Movie~2). 
{For this two-beam system, it is convenient to represent the configuration by the collective coordinate $x_1+x_2$, where $x_1+x_2<0$ indicates that the system is displaced more to the left, while $x_1+x_2>0$ indicates a displacement more to the right. We emphasize, however, that }the presence of collective coordinates {is} not
required for our framework, and {is} merely convenient for this specific case.
Our data shows that the irreversible transition is associated with a 
{saddle-node, or}
fold bifurcation, and this is confirmed in our numerical model where we can follow both stable and unstable configurations (Fig.~\ref{fig:fold}b, Movie~2).

As we argue below, this situation exemplifies the central motif that causes path dependencies in multistable materials:
On one side of the 
fold {bifurcation}, there are two types of configurations (ancestor and sibling). Moving across the bifurcation, the ancestor vanishes, and the system then jumps to another configuration that we call the descendant. Note that while descendant and sibling are smoothly connected, they can be distinguished by the presence of the ancestor, or equivalently, their strain in comparison to the fold transition. {
In such a triplet, the ancestor can always be identified as the state to which one cannot return.}
{Hence, each crossing of a fold singularity is associated with three configurations. These }are connected by an irreversible (fold) transition from the ancestor to the descendant and a reversible transition between the descendant and sibling (Fig.~\ref{fig:fold}c-d, Movie~2). 

\section{GENERAL APPROACH FOR SEQUENTIAL VECTORIAL DRIVING}\label{sec:General}

{We now present a general framework to describe the response for arbitrary driving paths, based on the central role that 
fold singularities and ADS-triplets play, 
and illustrate our approach using sample A. }
To describe the response to {\em any} {generic} driving path, we
introduce strain maps and path-transition graphs (pt-graphs). Strain maps encode
the location of the fold singularities
in driving space, and pt-graphs describe transitions between states that occur when the driving path crosses one of these fold singularities. Hence, nodes and edges of the pt-graph describe
states and their transition, and the edges
are labeled by their respective fold singularity.
The edges either are directed, corresponding to irreversible transitions (such as the path from ancestor to descendant), or 
undirected, corresponding to 
reversible, smooth transitions (such as {the} path between
descendant and sibling). 
Constructing this description requires three steps:
first identify the strain map; second, use it to define
states, and third, determine the 
pt-graph --- {for more details}, see \cite{MeulblokNonGeneric}.

\begin{figure}[t]
\centering
\includegraphics[]{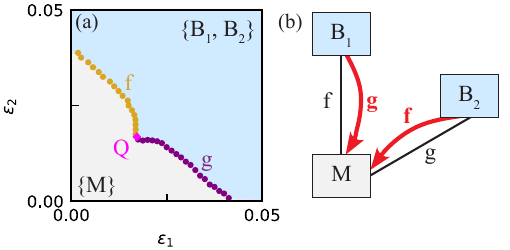}
\caption{
{The path-dependent response described by a strain-map (a) and pt-graph (b), here for sample A ($n=2, k\approx 0.09$).
(a) 
The locations of the fold transitions $f$ and $g$ 
form {curves of codimension one} that 
separate the strain space into a monostable domain with state $M$ (gray) and a bistable domain with states $B_1$ and $B_2$ (blue). The curves $f$ and $g$
meet in point $Q$ which corresponds to a {codimension-two} cusp singularity, or pitchfork bifurcation
(Appendix~C). {The error bars of the measured points in $f$ and $g$ are smaller than the symbol size.} 
(b) The pt-graph describes the smooth reversible connections (undirected black edges) and
discontinuous irreversible transitions (directed red edges)
for strain paths that cross the respective fold singularity curves of the strain map.
	}} \label{fig:ptgraph}
\end{figure}

We obtain the strain map for sample A
by determining
the locus in strain space of the fold singularities that govern the irreversible transitions
for $0\le \varepsilon_i \le 0.04$
(Fig.~\ref{fig:ptgraph}a).
To do so, 
we explore CCW and CW loop-like driving protocols, 
$P_5 : [c_1, c_2, d_1, d_2]$ and $P_6 : [c_2, c_1, d_2, d_1]$;
in $P_5$ ($P_6$) we fix $\varepsilon_{1}^M=0.04$
($\varepsilon_{2}^M=0.04$) and sweep $\varepsilon_{2}^M$ ($\varepsilon_{1}^M$).
Using $P_5$, we obtain the locus of the irreversible transitions ($f$) from 'left-leaning' to 'right-leaning' configurations; $P_6$ allows to define the irreversible transitions ($g$) from 'right-leaning' to 'left-leaning' configurations (Fig.~\ref{fig:ptgraph}a). The curves $f$ and $g$ thus specify the location of the fold singularities in strain space. For this specific example, $f$ and $g$ meet in a singular point $Q$ that corresponds to a cusp singularity, associated with a pitchfork bifurcation from a symmetric configuration where $x_1=-x_2$ to a pair of asymmetric configurations (Fig.~\ref{fig:ptgraph}a; appendix~C).

Second, for a given range of strains, the strain map 
partitions the strain space into domains. 
We define states as equivalence classes of configurations that are smoothly connected {\em within each domain}. For this example, 
{the loci of the fold bifurcations form
the fold curves $f$ and $g$, which}
partition the strain space into a monostable domain and a bistable domain, leading to three states
(Fig.~\ref{fig:ptgraph}a).
The evolution of the configuration in the monostable domain is smooth and reversible, and we thus associate all these configurations with a state $M$. In the bistable domain, we find two stable configurations for any value of the strain. These two configurations smoothly deform along strain paths within the bistable domain, defining two states $B_1$ and $B_2$ - note that
configurations associated with states $B_1$ and $B_2$ cannot smoothly deform into each other for strain paths within the bistable domain. 
In passing, we note
that our definition of states
as configurations smoothly connected within domains
explains why we distinguish between the descendant and sibling: these two states correspond to smoothly connected configurations, but in two different domains. 
Hence, 
the loci of the fold bifurcations {determine} a strain map and {define} the states; for example, the strain map has two fold curves
(Fig.~\ref{fig:ptgraph}) and three states
($M$, $B_1$ and $B_2$).

Third, to specify the full path-dependent response, we construct a pt-graph. The nodes of this mixed graph correspond to the states $M$, $B_1$ and $B_2$, and its
edges correspond to paths that cross one of the fold singularity curves, and are labeled accordingly.
The pt-graph combines directed and undirected edges.
The undirected edges correspond to smooth, reversible paths between descendant and sibling states that, while smoothly connected,
occur in different domains in strain space; the directed edges
correspond to 
non-smooth,
irreversible paths from an ancestor state to a descendant state.
Hence, to construct the pt-graph, we 
determine how states are connected when {\em crossing} the fold curves (in this paper, we ignore singular paths,
such as paths crossing through $Q$). For our specific example, the situation is simple. State $M$ smoothly transforms into state $B_1$ when we cross from the monostable to the bistable domain across $f$, and smoothly transforms into state $B_2$ across $g$; these smooth paths are reversible, so that $B_1$ and $B_2$ smoothly connect to $M$ across $f$ and $g$, respectively. 
The irreversible transitions occur when starting in the bistable domain. For instance, beginning in state $B_1$, crossing boundary $g$ leads to an irreversible transition to $M$ ($B_1 \rightarrow M$). Similarly, crossing boundary $f$ results in the irreversible transition $B_2 \rightarrow M$ (Fig.~\ref{fig:ptgraph}).

{This example illustrates general 
features of pt-graphs.
First, each fold curve is associated with an ancestor, descendant, sibling (ADS) triplet.
In particular, $B_2$, $M$, and $B_1$
are the ancestor, descendant, and sibling associated with curve $f$. 
Simiarly, $B_1$, $M$, and $B_2$ form the ADS-triplet of curve $g$. 
Second, the topology of the strain map constrains the number of edges in the pt-graph, as each state and each adjacent fold-boundary correspond to an outgoing edge in the pt-graph. Hence, the number of outgoing edges is $\Sigma_\mathrm{domain} n_s n_f $, where $n_s$ and $n_f$ are the number of states and adjacent fold curves per domain, respectively; here $\Sigma_\mathrm{domain} n_s n_f =6$. As reversible edges
in the pt-graph correspond to a pair of incoming and outgoing edges, this implies that
$\Sigma_\mathrm{domain} n_s n_f = 2n_r + n_i$, where
$n_r$ and $n_i$ are the numbers of reversible and irreversible edges of the pt-graph{, respectively}. We will discuss pt-graphs for more complex singularities elsewhere \cite{YangPNAS2023,OverveldePNAS2015,MeulblokNonGeneric}.}

The combination of the strain map and the pt-graph together captures the general features of the path-dependent response in multistable systems for {\em any} generic strain path (i.e., not crossing singular points such as $Q$).
This provides a unified framework to describe non-Abelian and mixed-mode responses, as well as transient and complex memory effects under vectorial sequential driving.
In particular, they allow to determine precisely when certain types of responses occur:
starting from any configuration associated with state $M$, a non-Abelian response occurs for
two paths that either cross $f$ or $g$ and meet in the bistable domain. Similarly, starting from any configuration associated with $B_1$ or $B_2$, a mixed-mode response occurs for pairs 
of strain paths that cross $f$ or $g$ and meet in the monostable domain.

{
The combination of a strain map and {a} pt-graph provides a complete description of the sequential response to vectorial driving, analogous to transition graphs (t-graphs) that describe the sequential response
under scalar driving \cite{RegevPRE2021,LindemanSciAdv2021,KeimSciAdv2021,LiuPNAS2024,MunganPRL2019,PaulsenPRSA2019,MelanconAdvFuncMat2022,JulesPRR2022,TeunisseArXiv2024,HeckePRE2021}. 
Determining such complete graphs can be challenging, in particular for systems with many states, and for high-dimensional driving. However, 
t-graphs have proven to be indispensable for understanding the sequential response of multistable systems subject to scalar driving \cite{RegevPRE2021,LindemanSciAdv2021,KeimSciAdv2021,LiuPNAS2024,MunganPRL2019,PaulsenPRSA2019,MelanconAdvFuncMat2022,JulesPRR2022,TeunisseArXiv2024,HeckePRE2021}, and we anticipate that pt-graphs will play a similar role for systems under vectorial driving. Moreover, as we will show below, 
the strain map and pt-graph provide a solid conceptual foundation on which to build simplified, effective descriptions. 
} 

\begin{figure*}[t]
  \centering
  \includegraphics[]{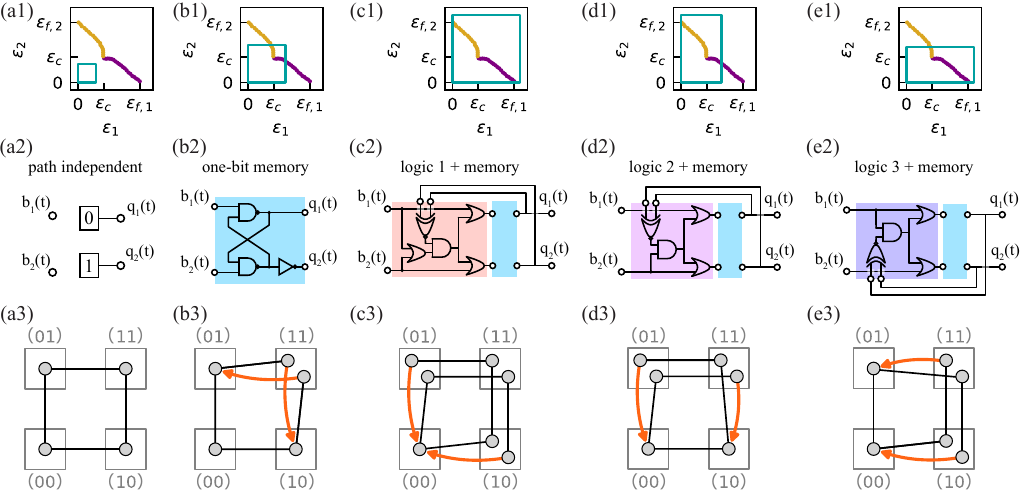}
  \caption{
  Sequential response of sample A to binary strain protocols.
  Top row: fold curves (yellow, purple) and binary strain protocols (green). For sample A, $\varepsilon_c\approx0.017$, $\varepsilon_g\approx0.018$, 
  $\varepsilon_{f,1}\approx0.039$ and $\varepsilon_{f,2}\approx0.041$. 
  In (a-c), we use symmetric strain scales $\varepsilon^M=0.01$, $0.025$, and $0.045$, respectively.
  In (d-e), the binary strain protocols are asymmetric with $(\varepsilon^M_1, \varepsilon^M_2)=(0.025, 0.045)$ and $(0.045, 0.025)$, respectively.
  Mid row: Corresponding Boolean circuits, where the blue blocks represent rs latches as in b2 (see appendix~D for details). 
  Bottom row: Corresponding b-graphs. Here, each node represents a configuration grouped (in squares) by the binary strain $b_i$ (labeled above or below the squares), and each edge {represents} a smooth (undirected and black) or irreversible (directed and orange) transition between two configurations.
  {See Movie~3 for the real-space evolution during each binary strain path and the construction of the corresponding b-graph.}
  }\label{fig:binary}
\end{figure*}

\section{{B-GRAPHS AND BOOLEAN CIRCUITS}} \label{sec:binary}

{
A complete characterization of the path dependency under vectorial driving of intermediate dimension generally leads to highly complex strain maps that, moreover, are difficult to obtain. 
To simplify the representation, we binarize the strains, i.e., take $\varepsilon_i=\varepsilon_i^{M} b_i$ where $\vec{b}=(b_1,b_2,\dots)$ and $b_i\in{0,1}$. 
We then consider strain trajectories composed of elementary steps, each changing a single $b_i$, and determine whether the resulting response is continuous or discontinuous.
This binarization of the strains substantially simplifies the analysis of vectorial path dependency. 
}

{
To describe the path-dependent response under binary strains, we introduce binary graphs ({\em b-graphs}). 
B-graphs provide a 
}
direct connection from the sequential response of multistable systems to {\em in materia} computing \cite{LiuPNAS2024}. 
We make this explicit by mapping the b-graphs of sample A onto Boolean logic circuits. 
Because the beam configurations are near-binary (left/right),
we project the real-space configuration $\vec{x}$ onto a binary output variable
$\vec{Q}=(q_1,\dots)$, where
$q_i=0$ (1) when $x_i<0$ ($x_i>0$). 
We stress that this binarization is merely a convenient feature of the beam system and is not required for the b-graph description, which only relies on identifying whether configurations are distinct. Nevertheless, the vector \( \vec{Q} \) enables a representation of the resulting sequential computations as Boolean logic circuits.

\subsection{{Boolean circuits}}
We determine the Boolean circuits that are embedded in Sample A. We study how the binary output
$\vec{Q}(\vec{b})$ responds to binary strain paths composed of compression steps ($c_i$) and decompression steps ($d_i$), starting from the neutral configuration $\vec{Q}=(01)$ at $\vec{b}=(00)$ (see appendix~D)
\footnote{For brevity of notation, we drop commas and write these binary quantities as $(00)$ instead of $(0,0)$}.

{ Different values of $\varepsilon_i^M$ can produce different
Boolean circuits. In sample A, there are three relevant strain scales:
at $\varepsilon_1=\varepsilon_2=\varepsilon_c\approx 0.017$ there is the cusp singularity; for 
$\varepsilon_i^M>\varepsilon_g\approx 0.018$, compressing a single beam changes the sign of $x_i$ of the other; and the fold curves
terminate at the axes for 
$\varepsilon=\varepsilon_{f,i}\approx 0.04$.
In the narrow interval
$\varepsilon_c<\varepsilon_i<\varepsilon_g$,
the binary projection is not suitable (see appendix.~C), and we focus on the pathways for $\varepsilon_i$ outside this interval.
}

Strikingly, we find five different sequential responses, depending on the strain scales $\varepsilon^M_i$ (Fig.~\ref{fig:binary}, Movie~3).
First, for small $\varepsilon_i^M< \varepsilon_c$, the strains are too small to reach the bistable domain and the response is smooth and path independent, yielding a trivial circuit (Fig.~\ref{fig:binary}a2).
For intermediate strains ($\varepsilon_g < \varepsilon_i^M < \varepsilon_{f,i}$), two stable configurations are reached for $\vec{b}=(11)$,
resulting in non-Abelian pathways and irreversible transitions (Fig.~\ref{fig:binary}b2, Movie~3). 
This path-dependent response is captured by {a} set-reset-latch (sr-latch), a simple circuit that can store one bit of information \cite{BooleanCircuits} (Fig.~\ref{fig:binary}b2; for details, see appendix~D).
For larger strains ($\varepsilon_i^{M} > \varepsilon_f$), every $\vec{b}\neq(00)$ produces two stable configurations, which can be represented by a
logic circuit preceding the same one-bit memory element (Fig.~\ref{fig:binary}c2, Movie~3).
Finally, asymmetric protocols
where one of $\varepsilon_1^M$ and $\varepsilon_2^M$
is in between $\varepsilon_g$ and $\varepsilon_f$, and the other is above $\varepsilon_f$ can be represented 
variations of such circuits (Fig.~\ref{fig:binary}d2-e2, Movie~3).
Thus, by selecting the magnitude of the binary strains at either `input', sample A can mimic four related yet distinct non-trivial Boolean circuits.

\begin{figure*}[t]
	\centering
	\includegraphics[]{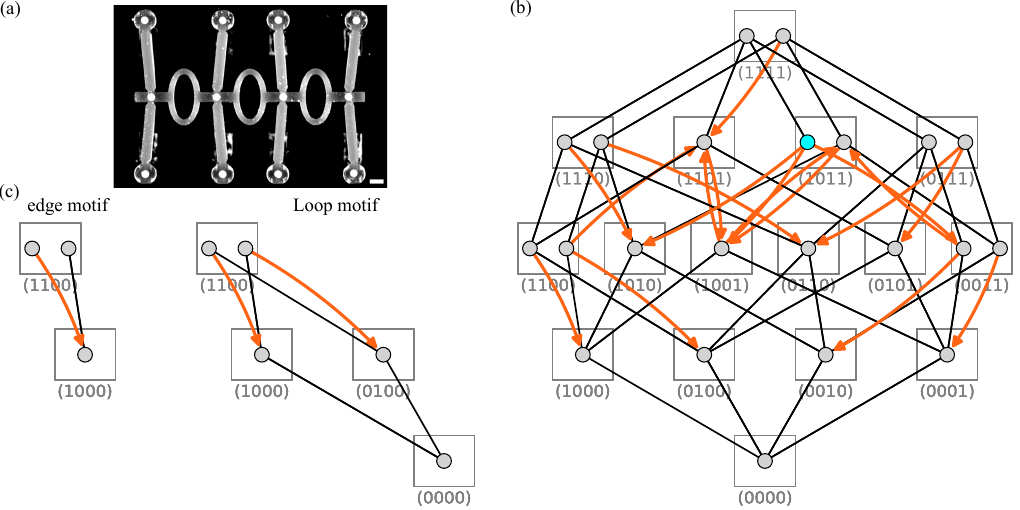}
	\caption{(a) Four-beam ($n=4$) sample B (scale bar is 5~mm, see appendix~A for details). (b)
  B-graph for sample B at $\varepsilon^M=0.035$. Each rectangle represents a binary strain $\vec{b}$ as labeled. 
  We organize all $2^4$ strain states by $\sum b_i$ along the vertical axis, and by their {lexicographical} order along the horizontal axis.
  The grey configurations are found by sequential compression of the rest-configuration at $\vec{b}=(0000)$; the 
  blue configuration  
  requires one decompression step, e.g., $d_2(1111)$.
  (c) Examples of low-dimensional building blocks (motifs) of the b-graph in (b). left: one-dimensional 'fold' (or 'ADS') edge motif; right: two-dimensional 'non-Abelian' loop motif. 
	}\label{fig:fourbeamresponse}
\end{figure*}

\subsection{B-graphs}

While the representation using Boolean circuits highlights the potential for {\em in materia} computing with multiple inputs, 
{it relies on a binarization of the system's configurations $\vec{x}\rightarrow\vec{Q}$, which is not always possible. Moreover, 
}
Boolean circuits are hard to interpret, in particular for larger systems, and technical issues arise from the frequent occurrence of race-conditions in such circuits --
{During race conditions, the logic state flips at multiple locations in the circuit simultaneously, leading to sensitivities to tiny differences in their timing. These race conditions are
a major reason for the widespread use of clocked, synchronous circuits in digital technology, rather than more frugal, faster, and energy-efficient asynchronous circuits.}

{We therefore introduce binary graphs (b-graphs), which retain the full output configuration descriptor $\vec{x}$ and capture the evolution to binarized strains
for a specific fixed set of strain scales $\varepsilon^M_i$
 (Fig.~\ref{fig:binary}a3-e3, Movie~3). In these b-graphs,}
each unique configuration corresponds to a node, and all nodes are grouped according to their respective binary strain $b_i$.
The smooth or non-smooth evolution 
during each (de)compression step is represented
by undirected and directed edges, respectively.
A b-graph is thus a mixed graph where each node has $n$ outgoing edges corresponding to the $n$ possible binary (de)compression steps, as we show for sample A (Fig.~\ref{fig:binary}a3-e3, Movie~3).

{The b-graph depends on the choice of $\varepsilon_i^M$ --- so that a single system described by one pt-graph/strain map may encompass multiple b-graphs (Fig.~\ref{fig:binary}).
B-graphs thus} contain a subset of the information in pt-graphs and circumvent the
challenges associated with obtaining complex strain maps, in particular for higher-dimensional strains.
{B-graphs thus
provide} a practical, scalable, and natural strategy to explore the salient features of the path-dependent response.
In particular, we note that the non-Abelian and mixed-mode response, as well as the ADS-pattern, can be represented in b-graphs.

\subsubsection{{High-dimensional strain space}} \label{sec:fourunit}
{For increasing number of units and
inputs, the strain space becomes high-dimensional, and keeping track of all domains and singularity boundaries in a pt-graph becomes challenging. 
To demonstrate that b-graphs allow to characterize 
the path-dependencies of systems with moderately high-dimensional
vectorial driving, we} determine the b-graph for a four-unit ($n=4$) sample B with the same geometrical parameters as sample A ($k=0.09$, $\delta=0.08$) at fixed $\varepsilon^M_i=\varepsilon^M=0.035$  
(Fig.~\ref{fig:fourbeamresponse}a-b).

Starting from the initial monostable configuration at $\vec{b}=(0000)$, we obtain 
all reachable configurations by a recursive strategy:
First, we apply all $4!$ compression sequences that reach $\vec{b}=(1111)$, yielding 
21 distinct configurations
{--- in the appendix~E we list the corresponding real space configurations}. 
Next, for each of these {states}, we apply any possible single-step decompression, which {in this example} yields one additional configuration.
We repeat this procedure until no new configurations are discovered, and keep track of the smooth or non-smooth nature of all transitions
 (see appendix~E).
This yields a b-graph consisting of $n_c=22$ configurations, $n_s=35$ smooth transitions, and $n_i=18$ irreversible transitions;
note that in general, $2\cdot n_s+n_r=n\cdot n_c$
(Fig.~\ref{fig:fourbeamresponse}b). {This b-graph provides a much simpler representation of the sequential response of this sample - albeit for fixed values of $\varepsilon^M_i$ - than could be achieved using pt-graph together with a four-dimensional strain map. }
We {also} have determined the b-graphs for sample B for other values of $\varepsilon^M$ experimentally (see appendix~E), and will discuss the full diversity of b-graphs that can be observed numerically elsewhere \cite{MeulblokMotifs}. 

{
We close by summarizing the main points. Binarized input variables allow the sequential response of multistable systems under vectorial driving to be represented by b-graphs, which are straightforward to obtain and much easier to visualize than pt-graphs and high-dimensional strain maps. They therefore provide a useful framework for describing the response in terms of sequential logic. For a single sample, varying the strain magnitudes yields multiple b-graphs, demonstrating that even moderately sized systems are computationally rich. More generally, our example illustrates that b-graphs are well suited to systems with a moderate number of driving dimensions.}

\section{{MOTIFS}} \label{sec:motifs}
{How can b-graphs be systematically analyzed, and how can we deal with systems with much larger driving dimensions? To address both questions, we introduce motifs {(Fig.~\ref{fig:fourbeamresponse}c)}. These can be seen as low-dimensional building blocks of b-graphs, but also offer a practical tool for characterizing the statistical response of very high-dimensional systems, where a full description of the response or b-graph is neither feasible nor particularly insightful.}

{We illustrate the notion of motifs first for the b-graph
 of sample B (Fig.~\ref{fig:fourbeamresponse}b).
 } Geometrically, binary strains
can be represented as vertices of an $n$-dimensional hypercube, where the $2^n$ (here: 16) vertices correspond to $\vec{b}$, the $n\cdot2^{n-1}$ (here: 32) edges represent the paths between adjacent strains, 
and the $n(n-1)\cdot2^{n-3}$ (here: 24) faces correspond to loops involving (de)compressing two strains. 
We focus here on 
{\em edge motifs}, which capture the qualitative patterns of configurations and transitions between adjacent strains, and
{\em loop motifs}, which describe these patterns associated with the two-dimensional faces {(Fig.~\ref{fig:fourbeamresponse}c)}.
We deal with symmetries by 
considering motifs equivalent
under flipping of the compression and decompression of each strain, i.e., $b_i \leftrightarrow (1-b_i)$,
and
under relabeling of the configurations at a given strain. 
As we show below, 
motifs are powerful tools for identifying the main features 
and distinct path dependencies encoded by the b-graph: for example, sample B exhibits three {edges} and 10 loop motifs that are absent in the two-beam system,
and loop motifs allow to detect the paradigmatic non-Abelian and mixed-mode path dependencies (Fig.~\ref{fig:fourbeamresponse}c).

\begin{figure}[tb]
	\centering
	\includegraphics[]{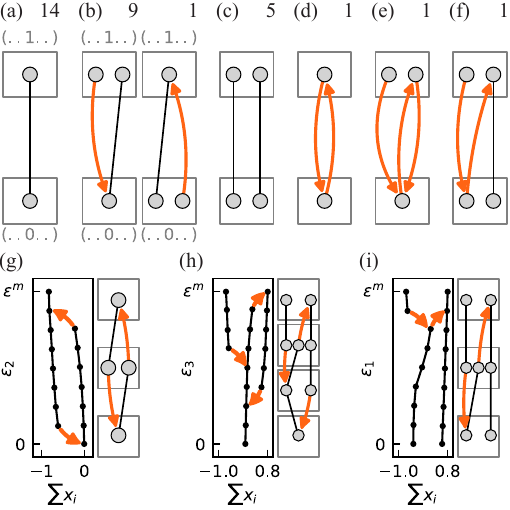}
	\caption{(a-f) The edge motifs of sample B at $\varepsilon^M=0.035$. The top-right number indicates how often these are observed {out of 32}, showing that smooth (a) and fold (b) motifs are most frequent. Note that while in (a-b) we indicate the binary strains $(\dots 0 \dots)$ $(\dots 1 \dots)$, we omit these in further panels as we do not distinguish between increasing and decreasing strains (i.e., the two examples in (b) are equivalent).
  (c) The double smooth motif is a juxtaposition of two smooth motifs, and thus reducible. (d-f) Three more complex, reducible edge motifs occurring once in the b-graph.
  (g-i) We follow the evolution of the corresponding configurations along the strain paths of motifs (d-f) and characterize the configurations by collective coordinate $\sum x_i$ (strain paths between $[b_i=[(1001), (1101)]$ (g), $[(1001), (1011)]$ (h), and $[(0011), (1011)]$ (i)). In each case, we find that the evolution along these paths combines two or three-fold motifs (left panels), so that we can compose the motif from more elementary smooth and fold motifs (right panels). Note that, although the two-fold singularities in (i) occur in the same experimental interval, these are independent events.
}\label{fig:edgemotifs_exp}
\end{figure}

\subsection{Edge motifs}
We determined the six edge-motifs of sample B,
three of which only occur once (Fig.~\ref{fig:edgemotifs_exp}a-f). We refer to the two simplest motifs as the smooth motif
and the fold motif (Fig.~\ref{fig:edgemotifs_exp}a-b). These motifs are dominant in many b-graphs and, in particular,
are the only edge motifs found in the two-beam sample A
(Fig.~\ref{fig:binary}). 
The smooth motif is characterized by a smooth, reversible path between
two single configurations and 
arises when there are no fold singularities along the strain path\footnote{
The reverse is not true: for example, the path between a descendant and sibling is also smooth even though it crosses a fold {curve}, but if the corresponding ancestor is not observed, this path cannot be distinguished from a simple smooth motif.
} (Fig.~\ref{fig:edgemotifs_exp}a).
The fold-motif contains a descendant for one strain 
and an ancestor-sibling pair for the other, and 
emerges when the path crosses a single fold singularity\footnote{
As before, the reverse is not necessarily true. 
} (Fig.~\ref{fig:edgemotifs_exp}b).
We note that
the descendant can be associated with either the
compressed or decompressed strain --- compare, e.g., the transitions between
$\vec{\varepsilon}_b=(1100)$ and $(1000)$ and between $\vec{\varepsilon}_b=(1100)$ and $(1101)$ (Fig.~\ref{fig:fourbeamresponse}) {--- and we
do not distinguish between these as
we considering motifs
equivalent under flipping of compression and decompression}
(Fig.~\ref{fig:edgemotifs_exp}b).

{The more complex edge-motifs are {\em reducible} -- composed of multiple smooth and fold motifs via juxtaposition and/or concatenation (Fig.~\ref{fig:edgemotifs_exp}g-i).}
The simplest example of a reducible motif is a juxtaposition of motifs, such as the double smooth motif, that can be broken up into two disjointed smooth motifs (Fig.~\ref{fig:edgemotifs_exp}c).
The other motifs
(Fig.~\ref{fig:edgemotifs_exp}d-f) are associated with paths that cross multiple fold singularities
and can be formed by concatenation
 (Fig.~\ref{fig:edgemotifs_exp}g-i).
For instance, 
consider the pair of irreversible paths that is observed, \textit{e.g.}, on the path between $\vec{\varepsilon}_b=(1001)$ and $(1011)$ (Fig.~\ref{fig:edgemotifs_exp}d and
Fig.~\ref{fig:fourbeamresponse}).
Following the configuration along this path
reveals two fold singularities (Fig.~\ref{fig:edgemotifs_exp}g), so that 
this motif is composed of a pair of fold motifs.
Similarly, other complex motifs (Fig.~\ref{fig:edgemotifs_exp}e-f) are similarly 
reducible and composed from smooth and fold motifs
(Fig.~\ref{fig:edgemotifs_exp}h-i).
These examples illustrate that, as
generic paths only cross fold singularities; any edge motif observed is composed of
smooth and fold motifs, highlighting their fundamental role.

\begin{figure*}[tb]
  \centering
  \includegraphics[]{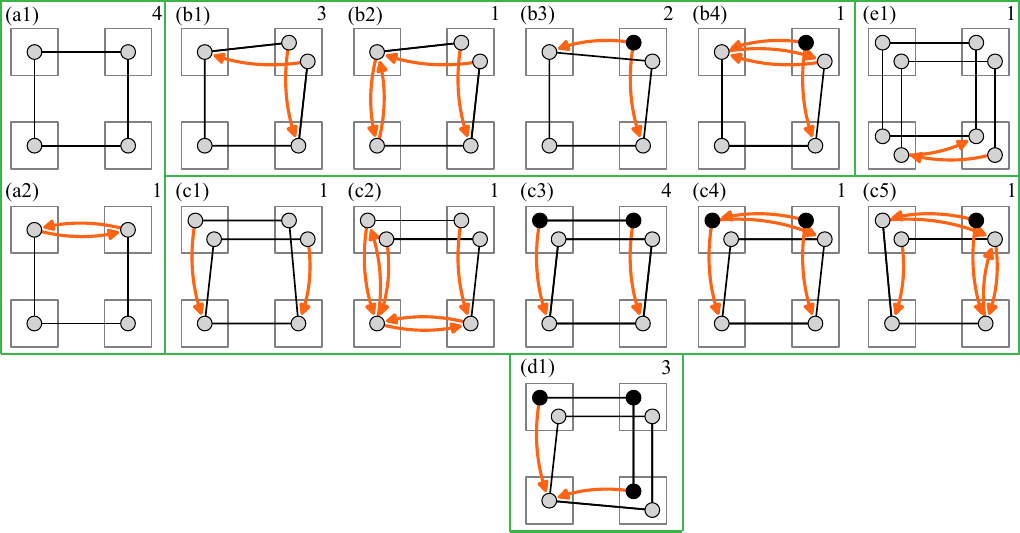}
  \caption{
The loop motifs of sample B at $\varepsilon^M=0.035$.
The black nodes lack an incoming connection in the motif, and correspond to 'no-return' configurations in their specific motif.
We group the motifs by their number of configurations, and orient these such that the larger number of configurations are to the right and top. Five of these motifs 
(a1,b1,b3,c3,d1) occur more than once in sample B, indicated by the top-right number. 
  }\label{fig:LoopMotifs_exp}
\end{figure*}

\subsection{Loop motifs}

We now consider loop motifs, which 
probe non-trivial path dependencies unique to vectorial strain inputs by 
involving two 'active' beams, while keeping all other strains constant.
As before, we consider two motifs equivalent under permutations of the two active beams and
flipping of compressions and decompressions.
By extracting all loop motifs from the b-graph of sample B, we identify {13} distinct loop motifs, which include those observed for two-beam sample A (Fig.~\ref{fig:LoopMotifs_exp}).
{Note that six of these motifs feature no-return configurations. 
These configuration feature no incoming transitions within the loop, and can only be reached by strain protocols involving the passive beams (Fig.~\ref{fig:LoopMotifs_exp}).}
We stress that each of these {represents} a distinct sequential logical operation, and discuss the emergent computational power of this system in section \ref{sec:six}.

While all edge motifs can be constructed from just two 
irreducible motifs - smooth and fold motifs - the situation
for loop motifs is significantly more complex. To identify the irreducible loop motifs, we focus on those that persist when the size of the loop in strain space is shrunk, 
and consider all different arrangements of fold curves
that cross the four edges composing a loop.
Apart from the trivial loop where all edges are smooth, there are three generic possibilities for the arrangements of the fold curves:
either the loop is intersected by a single fold curve, the loop contains two fold curves that intersect within the loop, or the loop contains two fold curves that terminate in a higher order cusp singularity.
All irreducible loop motifs can then be constructed starting from the combinatorics of the
fold curves intersecting different edges --- however, this construction is not trivial. Without going into too much detail, we note that fold curves have an orientation
(on which side lies the ancestor node), 
cusp structures can correspond to both supercritical and subcritical bifurcation, which lead to different motifs, and 
irreversible transitions can connect to multiple different descendants. We estimate that there are significantly more than 20 of such irreducible loop motifs.

\section{EMERGENT MEMORY AND COMPUTING}\label{sec:six}

We now discuss the path-dependent response under 
vectorial driving from two perspectives: as an emergent phenomenon that leads to novel memory effects that 
allows to understand and classify disordered media (Sec.~\ref{sec:tau}),
and as a new avenue towards {\em in materia} computing (Sec.~\ref{sec:tuncomp}).

\subsection{Transient responses} \label{sec:tau}

The response to cyclic driving provides an important fingerprint of 
the memory effects of complex systems \cite{PaulsenPRL2014,AdhikariEPJ2018,KeimPRR2020,BensePNAS2021,ShohatPNAS2022,KeimRMP2019,PaulsenAR2025,LindemanSciAdv2021,RegevPRE2013,SchreckPRE2013,RoyerPNAS2015,KawasakiPRE2016,LavrentovichPRE2017,NagasawaSM2019,YehPRL2020,KwakernaakPRL2023,KeimSciAdv2021,LiuPNAS2024,MunganAHP2019,MunganPRL2019,PaulsenPRSA2019,TerziPRE2020,DingJCP2022,LindemanSciAdv2025,KeimPRL2011,HaghPNAS2022}.
For scalar cyclic driving, deterministic, dissipative, multistable systems must settle in a periodic response after a finite number of cycles
\cite{BensePNAS2021,ShohatPNAS2022,KeimRMP2019,PaulsenAR2025,LindemanSciAdv2021,RegevPRE2013,SchreckPRE2013,RoyerPNAS2015,KawasakiPRE2016,LavrentovichPRE2017,NagasawaSM2019,YehPRL2020,KwakernaakPRL2023,KeimSciAdv2021,LiuPNAS2024,PaulsenPRSA2019}. For a wide class of systems that can be seen as composed of weakly interacting elements, the response satisfies so-called Return Point Memory which forbids the presence of transients longer than
one cycle
\cite{SenthaPRL1993,TerziPRE2020}.
Recent observations of finite transients, in, e.g., disordered media and metamaterials, are thus a hallmark of complex memory \cite{BensePNAS2021,ShohatPNAS2022,KeimRMP2019,PaulsenAR2025,LindemanSciAdv2021,RegevPRE2013,SchreckPRE2013,RoyerPNAS2015,KawasakiPRE2016,LavrentovichPRE2017,NagasawaSM2019,YehPRL2020,KwakernaakPRL2023,KeimSciAdv2021,LiuPNAS2024,PaulsenPRSA2019}. 

Here we analyze the edge and loop motifs of sample B from the perspective of transient responses. We first discuss 
the transients captured in edge-motifs, which are 
equivalent to the transients under scalar driving, and
then introduce loop-transients. We find that edge and loop transients can occur independently, and that the presence of loop transients may depend on the orientation of the driving loops. 
Vectorial or multi-input driving thus opens a new window on a wider diversity of 
complex memory effects.

First, edge motifs capture cyclic driving of 
a strain $\varepsilon_k$, and 
yield a sequence of configurations $c_{0} \rightarrow c_{1} \rightarrow c_2 \dots$ that 
must repeat. The index of the first repeating configuration, $t_e$,
defines the {\em transient time} $\tau_e:=t_e/2$ - for example, for a sequence of configurations
$\Gamma_0\rightarrow\Gamma_1\rightarrow\Gamma_2\rightarrow\Gamma_1\rightarrow\dots$, $\tau_e=1/2$
\cite{BensePNAS2021,ShohatPNAS2022,KeimRMP2019,PaulsenAR2025,LindemanSciAdv2021,RegevPRE2013,SchreckPRE2013,RoyerPNAS2015,KawasakiPRE2016,LavrentovichPRE2017,NagasawaSM2019,YehPRL2020,KwakernaakPRL2023,KeimSciAdv2021,LiuPNAS2024,PaulsenPRSA2019}. 
Clearly, the smooth motif leads to $\tau_e=0$ and the fold motif can produce $\tau_e=1/2$, provided the initial configuration is the ancestor (Fig.~\ref{fig:edgemotifs_exp}b). 
In composite motifs, which involve two or more folds, in our examples
$\tau_e=0,1/2$, and 1 respectively
(Fig.~\ref{fig:edgemotifs_exp}d-f).

\begin{figure}[t]
  \centering
  \includegraphics[]{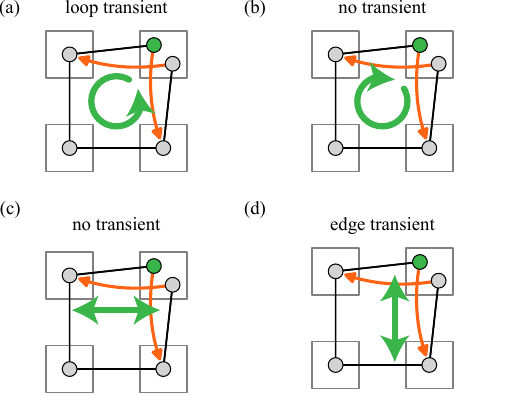}
  \caption{Illustration of the impact of orientation on the transient time. We focus on a single motif, and transient times under scalar and loop actuation (green arrow) starting from a given configuration (green circle).
  (a-b) The transient response of a loop-cyclic strain path depends on the orientation of the loop: in (a), {$\tau_l=1/4$}; in (b) $\tau_l=0$.
  (c-d) The transient response of an edge-cyclic strain path depends on the edge; (in (c), $\tau_e=0$; in (d) $\tau_e=1/2$).
    }\label{fig:FundamentalLoops}
\end{figure}

Next, we turn to loop motifs and the 
response to loop-cyclic strain paths that involve
two 'active' beams --- notice that such loop-cyclic paths can be
clockwise or counterclockwise.
We define the loop transient time as
$\tau_l:=t_l/4$, where $t_l>0$ is 
the index of the first repeating configuration.
{Except for} two very simple loop-motifs (Fig.~\ref{fig:LoopMotifs_exp}a), all others exhibit non-zero loop transients.
Starting from a no-return configuration, $\tau_l \ge 1/4$, and the longest transient observed 
is $(\tau_l=1)$ 
(Fig.~\ref{fig:LoopMotifs_exp}e{1}). 

Loop-transients exhibit properties that have no counterparts
in transients under scalar driving, which we illustrate using the cusp motif (Fig.~\ref{fig:LoopMotifs_exp}b1, Fig.~\ref{fig:FundamentalLoops}a-b).
The loop transient $\tau_l$ often depends on the loop-orientation{:
starting from the ``green'' configuration of the cusp motif, $\tau_l$ is 
1/4 for a CCW loop, and 0 for a CW loop (Fig.~\ref{fig:FundamentalLoops}a-b). T}his example illustrates that a non-zero loop-transient can occur along a totally smooth and reversible path --- something that is impossible along linear paths (Fig.~\ref{fig:FundamentalLoops}a).
Finally, the presence of loop and edge transients is 
decoupled.
Starting from the top node in the bistable region,
for the CCW loop-cyclic path $\tau_l=1/4$ (Fig.~\ref{fig:FundamentalLoops}a),
while an edge-cyclic path starting in the same direction has $\tau_e=0$ (Fig.~\ref{fig:FundamentalLoops}c)--- similarly, 
the CW loop-cyclic path leads to $\tau_l=0$ (Fig.~\ref{fig:FundamentalLoops}b), while the corresponding edge-cyclic path yields $\tau_e=1/2$ (Fig.~\ref{fig:FundamentalLoops}d).

\subsection{Tunable emergent computation} \label{sec:tuncomp}

\begin{figure}[tb]
  \centering
  \includegraphics[]{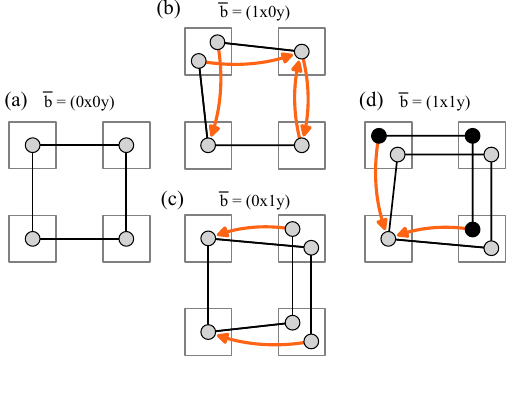} 
  \caption{
  Loop motifs of sample B at $\varepsilon^M=0.035$ with active beams 2 and 4 and distinct tuning beams. In this representation, the strains on the active beams 1 and 3 correspond to the $x$ and $y$ direction in the motifs, and the binary strains on the passive beams $b_1,b_3$ is 0 or 1, as indicated by the labels.
    }\label{fig:TunableComputing}
\end{figure}

The complex response of multistable materials with multiple inputs
opens a route to tunable computation. In our example of sample B, the loop motifs already uncover a wide range of different sequential logic circuits, but here we focus on the aspects of reprogrammability, where we use two inputs, e.g., 1 and 3, to tune the sequential logical operations on two other inputs, e.g., 2 and 4 (Fig.~\ref{fig:TunableComputing}). 
For this example, compressing the programming beams enables the selection of four distinct 
{mechanical algorithms}
that operate on the input beams.
Hence, high-dimensional input spaces allow for allocating specific input dimensions to tune the {\em in-materia} computing properties of the remaining input dimensions. While 
a similar strategy has been used to obtain reprogrammable 
logical circuits in disordered electronic materials with multiple terminals, we believe that our example illustrates that multistable systems naturally allow for sequential operations that mix logic and memory \cite{SerraGarciaPRE2019,ChenNat2020,ElHelouNat2022,KasparNat2021}. {In general, we expect that these computations can be captured by finite
state machines \cite{LiuPNAS2024}, where it is an open question how the number of states, input characters and computational capabilities grow with the dimensionality of the driving input.}

We point out that a wider variety of more powerful computations can be realized by using more complex driving protocols, samples with different interaction strength $k$, { and more complex geometries with more inputs.}
First, relaxing the constraints that the minimal input strains, \( \varepsilon_i^m \) are zero and the maximal input strains, \( \varepsilon_i^M \) are equal, naturally expands the space of computations and reprogrammabilities. Similarly, more information
can be encoded using non-binary driving sequences, pushing the 
computational power of our multistable samples in the direction of finite state machines with many states and input characters \cite{LiuPNAS2024}.
Second, in numerical simulations, we have observed that for lower values of $k$, the number of stable configurations per strain can reach $2^n$, and the number of reachable configurations is at least five. Hence, such samples would allow for longer and more varied transients and multi-bit sequential operations.
Finally, our chain geometry encodes specific interactions, and planar or fully connected geometries may encode more complex computations.

\subsection{{Computational capabilities}}
{
To understand the computational capabilities of our systems, we note that there is a competition between the number of states that can be stored and the computational expressivity. This balance can be clearly illustrated using the beam system. 
In the limit where 
the horizontal springs have a vanishing spring constant $k_h\rightarrow0$,
a system with $n$ beams can, in principle, store $2^n$ distinct states. However, computational expressivity vanishes in this limit, since compressing the beams simply leads to independent and deterministic increases of the coordinates $x_i$.
In the opposite limit of a large horizontal spring constant, the system only has two stable states with constant $x_i-x_{i-1}$, thus the expressivity again vanishes. 
While there is no universal measure of sequential expressivity, and its relation to formal finite-state-machine descriptions \cite{LiuPNAS2024} or scaling with the number of beams \(n\) remains open, these limits suggest an optimum in design space where a large fraction of the \(2^n\) states is reachable and the final state depends intricately on the driving sequence.
 }

\section{{DISORDERED SYSTEMS}}\label{sec:disorder}

\begin{figure*}[htb]
  \centering
  \includegraphics[]{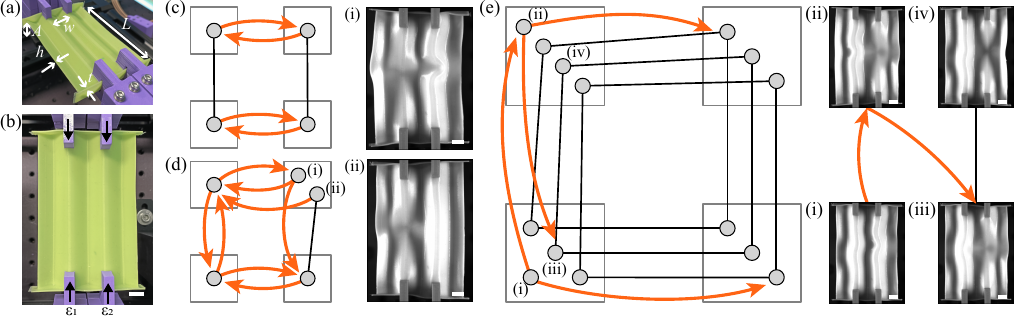}
  \caption{
  {Path-dependent response of disordered material formed by  
  a corrugated rubber sheet.
  (a) Sideview of the corrugated sheet clamped in our experimental device (for details see appendix~F).
  (b) Topview of the corrugated sheet compressed 
  by a vectorial drive $\vec{\varepsilon}=(\varepsilon_1,\varepsilon_2)$
  applied along the grooves at distinct locations as indicated. 
  Scale bars in (b-e) represent $10\,\mathrm{mm}$.
  (c-e) Loop motifs from binarized driving $\varepsilon_i= \varepsilon^m+ b_i\varepsilon^\Delta_i$, where $b_i=0,1$.  
  (c) Abelian loop motif with four irreversible transitions, for 
  $(\varepsilon^m, \varepsilon^\Delta_1, \varepsilon^\Delta_2)=(0,0.045,0.020)$.
  (d) A non-Abelian motif with two distinct configurations (i)-(ii) at $\vec{b}=(11)$ as shown on the right. Here
$(\varepsilon^m,\varepsilon^\Delta_1,\varepsilon^\Delta_2)=(0, 0.045, 0.045)$.
  (e) Complex motif containing both edge transients and loop transients for
  $(\varepsilon^m,\varepsilon^\Delta_1,\varepsilon^\Delta_2)=(0.020, 0.025, 0.025)$. 
  Panels (i)-(iv) show the evolution of an edge transient under cyclic driving.
  Other edge transients with $\tau_e=0.5$, as well as loop transients $\tau_l=0.25,0.5$ also arise in this motif.} {See Movie 4 for the evolution of the real-space configuration of the non-Abelian response of motif (d) and the edge transient in motif (e).}
  }\label{fig:GroovySheet}
\end{figure*}

{
To highlight the general applicability of our approach, we consider the response to vectorial driving of 
thin elastic sheets with parallel corrugations, which form
a disordered system (Appendix~F). In earlier work, we tracked the sequences of snapping transitions under
compression along the direction of the grooves and
showed that such
''groovy sheets'' are multistable
(Fig.~\ref{fig:GroovySheet}a) \cite{BensePNAS2021, MeeussenNat2023}. 
Despite their regular appearance, the response of groovy sheets is comparable 
to those of disordered systems:
the precise locations and critical strain associated with the snapping events are set by fabrication imperfections and are thus frozen yet essentially random \cite{BensePNAS2021}.
In addition, and in contrast to our chain-like metamaterials, 
the different states of the sheet are not easily distinguished by a set of predetermined coordinates. 
Hence, groovy sheets allow to test our approach in the context of disordered systems. 
}

{
To explore the motifs of this disordered system, 
we apply two-dimensional vectorial compression $\vec{\varepsilon}=(\varepsilon_1, \varepsilon_2)$ to two separated locations along the corrugations (Fig.~\ref{fig:GroovySheet}b) --- note that in previous work compression was homogeneous~\cite{BensePNAS2021}. Despite the absence of simple coordinates that characterize its evolution, we can easily determine experimentally 
whether {\em (i)} variations of $\vec{\varepsilon}$ 
lead to smooth deformations or snappy and irreversible transitions,
and {\em(ii)} whether two configurations are distinct or not \cite{BensePNAS2021}. As discussed before, this is sufficient to determine b-graphs.
}

{
To explore the rich path dependence of these sheets, we first scan strain space to identify regions of interest (Appendix~F). From this survey, we highlight three motifs - not observed in the chain -
that emerge under binary protocols where $\varepsilon_i= \varepsilon^m + b_i \varepsilon^\Delta_i$  (see Sec.~\ref{sec:methods}).
We first illustrate two non-trivial motifs observed for $\varepsilon^m=0$ (Fig.~\ref{fig:GroovySheet}c-d, Movie~4).
The first motif is Abelian, but shows a combination of 
smooth, reversible deformations and non-smooth irreversible transitions (Fig.~\ref{fig:GroovySheet}c).
The second motif is non-Abelian, demonstrating that such path dependent responses are a general feature of multistable materials. This specific motif exhibits an intricate combination of mostly irreversible transitions not seen in the chain
(Fig.~\ref{fig:GroovySheet}d, Movie~4). 
For the third motif, we use 
pre-compression with strain $\varepsilon_m=0.025$, where the energy landscape is expected to be more complex and contains more states \cite{BensePNAS2021}. We observe a complex motif that exhibits both edge transients 
($\tau_e=0.5,1$ ) and 
loop transients $\tau_l=0.25,0.5$ (Fig.~\ref{fig:GroovySheet}e, Movie~4). 
}

{
We note that these measurements are not sufficient to determine the PT-graph --- even though a given strain path between two configurations may show an irreversible transition, we cannot rule out that there are other paths where these configurations are smoothly connected. 
In contrast, 
determining the full set
of b-graphs, lets say for each loop of the form
$\varepsilon_i= \varepsilon_i^m + b_i \varepsilon^\Delta_i$ is possible although 
tedious. By focusing on the motifs one effectively can
sample the complexity of the sequential response under vectorial driving.}

{
This example illustrates the broad applicability of our framework, not only to model systems such as our metamaterial chain but also to disordered materials that lack a simple parametrization of their states. Vectorial driving reveals new insights into their sequential response and poses experimental constraints on putative models of 
complex materials. In this context, motifs provide a particularly effective tool for characterizing complex behavior.}

\section{CONCLUSION AND OUTLOOK}
{Vectorial driving reveals a wealth of path-dependent responses and offers new insights into the complexity of frustrated materials.
We motivated this paper with broad questions about new phenomena and computational capabilities, the description of vectorial responses, and how to address the complexity of high-dimensional driving. We now summarize our findings. 
}

{
First, we observe striking new phenomena such as the non-Abelian response, mixed mode response, and orientation-dependent loop transients (Secs.~\ref{sec:PDs} and \ref{sec:motifs}). These allow even small systems to act as memory latches
which are programmed by the sequence of driving
(Sec.~\ref{sec:binary}); more general, they open a route to multi-input finite state machines \cite{LiuPNAS2024}.
}

{
Moreover, we offered three related frameworks for the description of the path-dependent sequential responses. At the most fundamental level, pt-graphs provide a complete description of the sequential response to any driving protocol and generalize t-graphs from scalar to vectorial driving. This description directly links the 
path-dependent response to the underlying structure of singularities.
We show that co-dimension one fold singularities underpin all generic path dependencies and lead to ADS-triplets that form the basis
of graph-based descriptions of the sequential response.
Pt-graphs differ from t-graphs in that their edges can be reversible or irreversible and are labeled by boundaries in parameter space rather than single critical thresholds \cite{RegevPRE2021,BensePNAS2021,LiuPNAS2024,MunganAHP2019,MunganPRL2019,TerziPRE2020,HeckePRE2021,MelanconAdvFuncMat2022,JulesPRR2022,DingJCP2022}. However, the two are closely related: restricting a pt-graph to a single one-dimensional path yields a t-graph, and 
pairs of ADS-triplets form elementary hysteresis loops (hysterons) \cite{RegevPRE2021,BensePNAS2021,ShohatPNAS2022,LindemanSciAdv2021,KeimSciAdv2021,LiuPNAS2024,MunganAHP2019,MunganPRL2019,TerziPRE2020,HeckePRE2021,MelanconAdvFuncMat2022,JulesPRR2022,DingJCP2022}. 
}

{
Obtaining a complete pt-graph is challenging for large systems with many states or high-dimensional driving spaces, compounding a problem already present for scalar driving, where determining the full t-graph may not be possible or practical \cite{RegevPRE2021,BensePNAS2021,ShohatPNAS2022,LindemanSciAdv2021,KeimSciAdv2021,LiuPNAS2024,MunganAHP2019,MunganPRL2019,TerziPRE2020,HeckePRE2021,MelanconAdvFuncMat2022,JulesPRR2022,DingJCP2022}. We offer two strategies to address the most salient features of the response to vectorial driving. B-graphs simplify the description by binarizing the driving input, and we demonstrate 
that b-graphs can be obtained for four-input systems. This approach
provides an effective, simplified driving strategy that reveals new memory effects and computational capabilities
(Sec.~\ref{sec:binary}).
For even larger systems, we introduce motifs as a general strategy to probe the sequence-dependent sequential response for any system, regardless of its size, and demonstrate their applicability for complex chains and for
a disordered system
(Secs.~\ref{sec:fourunit} and \ref{sec:disorder}). 
}

{More broadly,
the historic increase in the complexity of measured data in physical experiments---from scalar measurements to images, and from static to dynamic data---is increasingly matched by a comparable increase in the complexity of driving protocols.
For example, several recent studies have explored multistable systems driven along multiple directions, including shear applied in different orientations
\cite{DagoisBohyPRL2012,GoodrichPRE2014,GendelmanEL2015,PatinetPRL2016,BarbotPRE2018,SchwenSM2020,LindemanArXiv2024,LindemanSciAdv2025},
uniaxial compression applied along different axes \cite{FlorijnPRL2014}, 
and sequential combinations of compression and shear \cite{OhernPRE2003,HeckeJPCM2010,LiuAR2010}.
While these works typically focus on specific protocols or responses, they nonetheless provide concrete settings in which vectorial driving naturally arises and complex path-dependent responses, including non-Abelian responses, emerge.
Our work differs in that we explicitly consider spatially textured driving and provide a framework---based on pt-graphs, b-graphs, or motifs---to systematically organize the resulting sequential responses. In addition, recent
work on rate-dependent responses naturally introduces a vectorial driving space and opens the door to systematic studies of memory effects under time-ordered inputs without requiring new experimental detection techniques \cite{GutierrezMeulblokArXiv2025,WatkinsArXiv2025,MonneryArXiv2025}.
More generally, any system with multiple controllable driving ``directions'' offers a natural platform in which the concepts developed here could be applied.
While our paper focuses on mechanical systems, we believe our approach may be relevant for a wide range of platforms, including in optics
\cite{Yang2019,Slim2024}, chemistry \cite{vanSluijsNatComm2022,McMullen2022} and electronics \cite{ChenNat2020}.
Together, vectorial driving opens new avenues for characterizing frustrated systems, uncovering novel multi-dimensional responses and memories, and advancing strategies for powerful {\em in-materia} computing \cite{KasparNat2021}. 
}

Finally, vectorial driving allows a new window on complex materials{, as illustrated in section \ref{sec:disorder}}.
{Even though the internal state of such materials may be complex \cite{CubukPRL2015}, as
long as one can determine the discontinuities along driving paths and can compare whether two configurations are equal, our framework is applicable.
One accessible and novel way to characterize extended disordered systems is to test whether their response to two-dimensional driving is non-Abelian as a function of the distance between driving locations and the driving strength, which may reveal characteristic length and actuation scales \cite{MeulblokMotifs}. }
Moreover, we expect 
a wealth of new memory effects, such as
multiperiodic responses under loop-like driving paths \cite{PaulsenPRL2014,BensePNAS2021,KeimRMP2019,PaulsenAR2025,KeimSciAdv2021,RegevPRE2013,SchreckPRE2013,RoyerPNAS2015,KawasakiPRE2016,LavrentovichPRE2017,NagasawaSM2019,YehPRL2020}, and higher-dimensional equivalences of return-point memory \cite{BensePNAS2021,ShohatPNAS2022,KeimRMP2019,PaulsenAR2025,MunganAHP2019,MunganPRL2019,TerziPRE2020}.

\section*{ACKNOWLEDGEMENTS}
CM and MvH acknowledge funding from European Research Council Grant ERC-101019474.
We thank M. Teunisse, Y. Shokef {and our referees} for fruitful discussions {and suggestions,} and J. Mesman, D. Ursem, H.J. Boluijt, and R. Zwart for technical support.

~\newpage
~

\cleardoublepage

\appendix

\section*{APPENDIX A: EXPERIMENTAL DETAILS} \label{sec:expdetails}

\subsection{Geometry and Fabrication}

Our samples A and B are fabricated by pouring degassed
Mold Star 30 silicone rubber
(Young's modulus of $\approx1$ MPa, Poisson's ratio of $\approx 0.5$)
into open face molds that are 3D printed on UltiMaker S3 printers.
The samples are cured at room temperature. After curing, the samples are removed by breaking the molds carefully.

\begin{figure}[ht]
	\centering
	\includegraphics[]{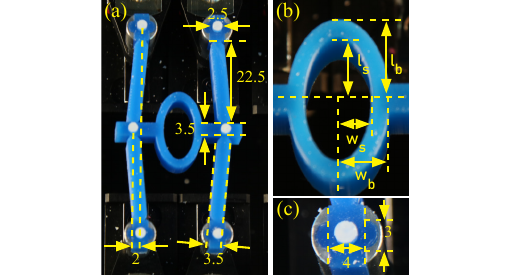}
	\caption{Sample geometry; all dimensions in milimeters, errorbar  $\pm 0.1~\text{mm}$. (a) Sample A with two units,
  (b) Zoom in on elliptical annulus. (c) Zoom in on terminating block.
	}\label{fig:geometry}
\end{figure}

Sample A features two units and sample B features four units; both have the same beam parameters ($t=3.5$ mm, $\delta=0.08$), 
a thickness of $10~\text{mm}$ to prevent out-of-plane buckling, and tapered connections to minimize torsional effects (Fig.~\ref{fig:geometry}a).
We embedded an elliptical annulus 
($w_b=12$ mm, $l_b =20$ mm, $w_s=8$ mm and $l_s=16$ mm) in the horizontal beam to lower its spring constant (Fig.~\ref{fig:geometry}b). 
Each unit is terminated by a rectangular block that allows to attach the sample to the compression device (Fig.~\ref{fig:geometry}c).
Finally, we use $2.5~\text{mm}$ diameter protrusions 
that are painted white
to track the real-space configuration (Sec.~\ref{sec:realspace}). 

\subsection{Measuring Effective Coupling}\label{sec:kstar}
We measured the mechanical response of the vertical beam and horizontal spring to determine the effective
interaction coefficient
$k=k_h/k_v$ of our samples.
We fabricate separated vertical units and elliptical spring units using the same 3D-printing and molding techniques, and probe their force-compression
response $F(U)$ in an uniaxial testing device (Instron type 5965), which controls the compression $U$ better than $10^{-3}$ mm and allows us to measure the compressive force $F$ with an accuracy $10^{-4}$ N. 
We use linear fits to $F(U)$ and find that $k_v=9.17\pm0.01 \cdot 10^2~\mathrm{N/m}$ and $k_h=81\pm5 ~\mathrm{N/m}$, leading to $k\approx0.09$.

\begin{figure}[hbt]
  \centering
  \includegraphics[]{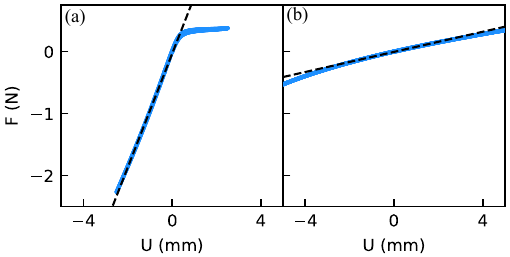}
  \caption{Mechanical response of a vertical beam (a) and horizontal spring (b) and linear fits (dashed lines). The vertical spring buckles at large compression ($U>0.7~\text{mm}$) - in our path dependent experiments, we do not reach this limit - and we restrict
the fit to the domain $U\in[-2~\text{mm},~0.5~\text{mm}]$ - 
  The dashed lines represent linear fits.
  }\label{fig:kstar}
\end{figure}

\subsection{Driving} \label{sec:realspace}

We use a custom-made compression device to individually control the compression on each beam.
The beams are symmetrically compressed from both top and bottom to ensure horizontal alignment of the middle nodes. 
The compression is controlled by a DC servo motor actuator (Thorlabs Z825B), providing an accuracy of $0.01$ mm and a repeatability of $0.01$ mm.
To maintain operation within the quasi-static limit, the compression rate is set to $0.1~\mathrm{mm/s}$, with a maximal acceleration of $3~\mathrm{mm/s^2}$.
During each compression sequence, 
snapshots of the metamaterial configuration are captured using a CCD camera (Basler acA2040-90um, $12.9$ pixel/mm). From the snapshots, the node positions are extracted using standard python image analysis packages, achieving an accuracy of $0.005$ of the dimensionless quantities 
 $\vec{x}$.

\subsection{{Repeatability}}
{
All strain protocols for each experimental b-graph shown in Figs.~\ref{fig:binary}, \ref{fig:fourbeamresponse}, and \ref{fig:GroovySheet} 
(The data of 
Figs.~\ref{fig:edgemotifs_exp}-\ref{fig:TunableComputing} are derived from these figures)
were repeated at least three times. 
In all cases, the measured responses were consistent across the repetitions. 
}

\section*{APPENDIX B: NUMERICAL DETAILS} \label{sec:numdetails}
Complementary to experiments, we use a numerical spring model.
We define the elastic energy $E(x_i)$ 
by approximating
each biased beam with two (nearly) vertical springs with spring constants $k_v$ and rest length $l_0=\sqrt{1+\delta^2}$, where $\delta$ is the bias of the beam.
The energy of each vertical spring reads:
\begin{align}
	E_{v_i} = \frac{1}{2}(l-l_0)^2 = \frac{1}{2} \left( \sqrt{ (1-\varepsilon_i)^2 + x_i^2 } - \sqrt{1+\delta^2}\right)^2~,
\end{align}
where, $x_i$ is the position of the middle of the beam with respect to the clamped-clamped boundaries and $2\varepsilon_i$ is the compressive strain.
The horizontal spring has alternating rest length $1-(-1)^i2\delta$ and spring constant $k_h$ and energy:
\begin{align}
	E_{h,i} = \frac{k}{2}(x_i - x_{i+1} - 2\delta)^2 ~.
\end{align}
As each unit in our system consists of two vertical springs, we obtain a total energy:
\begin{multline}
	E = \sum_{i=1}^{n} \left( \sqrt{ (1-\varepsilon_i)^2 + x_i^2 } - \sqrt{1+\delta^2} \right)^2 + \dots \\
	\sum_{i=1}^{n-1} \frac{k}{2} \left( x_i - x_{i+1} - (-1)^i2\delta) \right)^2 ~.
\end{multline}
where $k:=k_h/k_v$ is the ratio between the horizontal and vertical spring constants.

We obtain equilibrium positions of $\{x_i\}$ given a strain $\vec{\varepsilon}$ by finding the roots of the forces $F_i=\partial_{x_i}E$ using the Newton-Raphson method.
Each strain-driving step is divided into several small steps of $10^{-5}$ to mimic the quasi-static dynamics and to allow to distinguish between smooth and non-smooth deformations.

In the lowest order, the energy simplifies to:
\begin{equation}
	E = \sum_{i=1}^{n} \left( \frac{x_i^2}{2} - \gamma_i \right)^2 + \sum_{i=1}^{n-1} \frac{k}{2} \left( x_i - x_{i+1} - (-1)^i2\delta) \right)^2 ~,
\end{equation}
where $\gamma_i := \varepsilon_i + \delta^2/2 - \varepsilon_i^2/2 \approx \varepsilon_i + \delta^2/2$.
Hence, our model system essentially consists of coupled quartic potentials. 
In the small strain limit ($\varepsilon_i^2\approx0$), this approximation of $E$ can be rescaled such that the bias $\delta$ drops out:
\begin{align}
	x_i &\rightarrow \delta x_i'~, \\
	\varepsilon_i &\rightarrow \delta^2 \varepsilon_i'~, \\
	k &\rightarrow \delta^2 k~, \\	
	E &\rightarrow \delta^4 E'~,
\end{align}
resulting in:
\begin{multline}
	E' = \sum_{i=1}^{n} \left( \frac{x_i'^2}{2} - \varepsilon_i - \frac{1}{2} \right)^2 + \\ \sum_{i=1}^{n-1} \frac{k}{2} \left( x_i - x_{i+1} - (-1)^i2) \right)^2 ~.
	\label{eq:energylowestorder}
\end{multline}
Hence, in lowest order, the coupling strength $k$ is the only system parameter.

\section*{APPENDIX C: CONFIGURATIONS NEAR CUSP} \label{sec:Configs}
\begin{figure}[bt]
  \centering
  \includegraphics[]{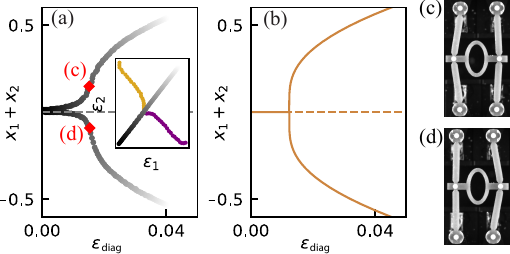}
  \caption{
  Pitchfork bifurcation observed along a non-generic strain path through a cusp singularity. (a) The collective coordinate $x_1 + x_2$ along a diagonal strain path ($\varepsilon_1=\varepsilon_2:=\varepsilon_{\text{diag}}$, inset)
  starting from $\varepsilon_{\text{diag}}=0.04$ and ending at $\varepsilon_{\text{diag}}=0$ (sample A).
  (b) Results from spring model ($k=0.09$ and $\delta=0.08$) along the same path. (c-d) Snapshots of the system corresponding to the red dots indicated in panel (a); note that in both cases, the left beam is left leaning and the right beam is right leaning.
  }\label{fig:pitchfork}
\end{figure}

Here, we detail the real-space configurations in the bistable domain near the cusp/pitchfork $Q$. 
So far, 
we have associated the non-Abelian response with 
$B_1$ and $B_2$ being 'left-leaning' or 'right-leaning' for sample A (see Sec.~\ref{sec:PDs}).
However, near the pitchfork bifurcation $Q$, 
the configurations associated with $B_1$ and $B_2$ both have
$x_1<0$ and $x_2>0$, and are thus not 'left-leaning' or 'right-leaning' (Fig.~\ref{fig:pitchfork}).
As a result, the binary projection of $\vec{x}$ does not capture the bistability in this (small) domain (Fig.~\ref{fig:pitchfork}a).

\section*{APPENDIX D: BOOLEAN LOGIC DETAILS} \label{sec:BooleanDetails}

\begin{figure*}[tb]
  \centering
  \includegraphics[]{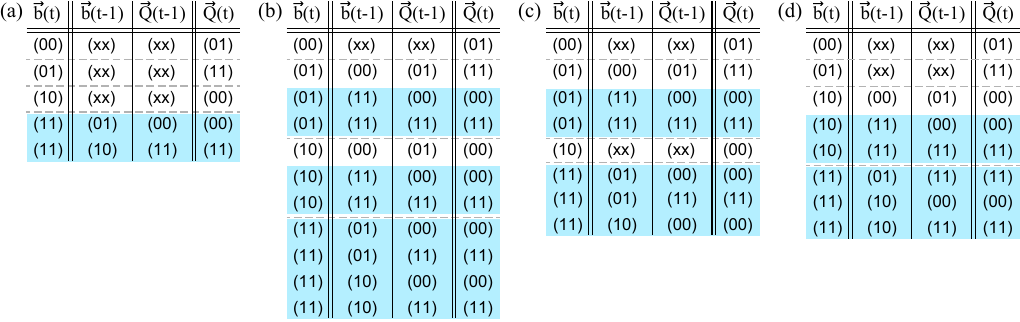}
  \caption{ 
  (a-d) Truth tables of sample A corresponding to the sequential Boolean circuits in Fig.~\ref{fig:binary}b-e, respectively; the table for Fig.~\ref{fig:binary}a is trivial and omitted here. Crosses indicate cases where the values of $\vec{b}(t-1)$ and $\vec{Q}(t-1)$ are irrelevant; blue boxes indicate cases where the output depends on the previous state.
  }\label{fig:truthtables}
\end{figure*}

Here, we detail the mapping from the five b-graphs, present in sample A, to their Boolean circuits.
{Recall that for this mapping} we project the 
real-space configuration onto a binary output variable $\vec{Q}=(q_1q_2)$, where $q_i=0$ ($q_i=1$) when $x_i<0$ ($x_i>0$). 
{Furthermore, we introduced}
a binary input variable $\vec{b}=(b_1b_2)$ with $b_i=0,1$, defining the strain as $\varepsilon_i=\varepsilon_i^Mb_i$, and
{consider sequential protocols where $\vec{b}(t)$ and $ \vec{b}(t-1)$ differ in precisely one location}.
As explained in the main text, we obtain the five different regimes by
varying $\varepsilon_i^M$, avoiding the tiny
interval $\varepsilon_c<\varepsilon_i^M <\varepsilon_g$ where the units are not clearly left or right leaning. 

To describe the five distinct path dependent response{s, we present} the relation between {$\vec{b}(t)$, $\vec{b}(t-1)$, $\vec{Q}(t)$ and $\vec{Q}(t-1)$} {in truth tables (Fig.~\ref{fig:truthtables}).}
{We note that the truth table of the linear regime ($\varepsilon^M_i<\varepsilon_c$) is trivial ($\vec{Q}(t)=(01)$ irrespective of (previous) inputs).}
In the four non-trivial truth tables, we highlight
cases where $\vec{Q}(t)$ depends on the previous state in blue (Fig.~\ref{fig:truthtables}).
In the main text, we map these
truth tables to logic circuits, although we stress that these circuits are not unique. Finally, 
we stress a subtle point between ordinary SR latches and our circuits: since the outputs of 
SR latches usually are supposed to be connected (i.e. $Q_1=\neg Q_2$, the input $\vec{b}=(00)$ is considered ill-defined - here {the outputs are independent, and therefore} all inputs are allowed.

\section*{APPENDIX E: EXPERIMENTAL FOUR-BEAM B-GRAPHS} \label{sec:exp-map-strat}
Here, we detail the four-beam b-graph of sample B and provide four additional b-graphs for distinct $\varepsilon^M$.
Different driving sequences that end in strain 
$\vec{b}$ can lead real-space configurations $\vec{x}$ that are only slightly different.
We take two configurations as equivalent if $\sum_i|x_i(P_k(\vec{b})) - x_i(P_l(\vec{b}))| \leq 0.015$.

{
Next, we provide a schematic state description of the b-graph in Fig.~\ref{fig:fourbeamresponse}b in Tab.~\ref{tab:statedescribtion}. To do so, we represent $x_i<0$ as $<$ and $x_i>0$ as $>$, thus each state can be presented by four symbols such as `$><><$'. 
}

\begin{table}[htb]
  \centering
  \begin{tabular}{|c|c|}
    \hline
    $\vec{b}$ & $\vec{x}$ \\ \hline \hline
    $(0000)$ & $(><><)$ \\ \hline
    
    $(1000)$ & $(>>><)$ \\ 
    $(0100)$ & $(<<><)$ \\ 
    $(0010)$ & $(><>>)$ \\ 
    $(0001)$ & $(><><)$ \\ \hline

    $(1100)$ & $(<<<<)$ \\ 
    $(1100)$ & $(>>><)$ \\ 
    $(1010)$ & $(>>>>)$ \\ 
    $(1001)$ & $(><><)$ \\ 
    $(0110)$ & $(<<<<)$ \\ 
    $(0101)$ & $(<<<<)$ \\ 
    $(0011)$ & $(><<<)$ \\ 
    $(0011)$ & $(>>>>)$ \\ \hline
    
    $(1110)$ & $(<<<<)$ \\
    $(1110)$ & $(>>>>)$ \\ 
    $(1101)$ & $(<<<<)$ \\ 
    $(1011)$ & $(<<<<)$ \\ 
    $(1011)$ & $(>>>>)$ \\ 
    $(0111)$ & $(<<<<)$ \\ 
    $(0111)$ & $(>>>>)$ \\ \hline
    
    $(1111)$ & $(<<<<)$ \\ 
    $(1111)$ & $(>>>>)$ \\ \hline
  \end{tabular}
  \caption{{States in the b-graph in Fig.~\ref{fig:fourbeamresponse}b, where $x_i<0$ is represented as $<$ and $x_i>0$ as $>$.}}
  \label{tab:statedescribtion}
\end{table}

\begin{figure*}[bt]
  \centering
  \includegraphics[]{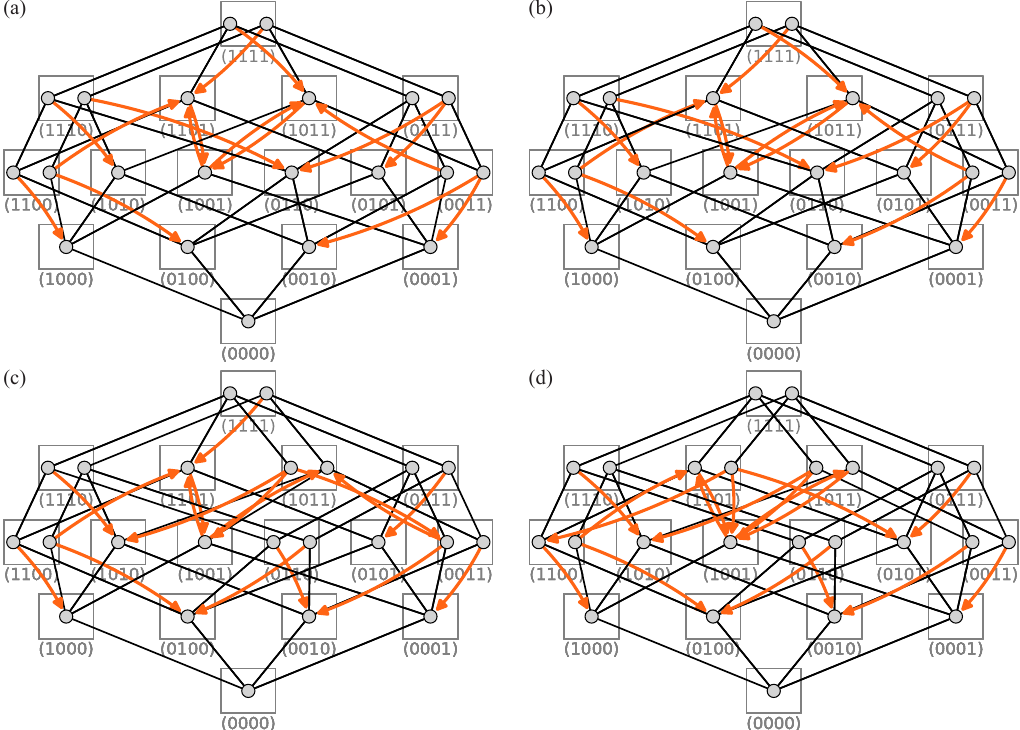}
  \caption{
  (a-d) b-graphs of sample B for $\varepsilon^M=0.03$, $0.0325$, $0.0375$, and $0.04$ respectively. We note that the b-graph for $\varepsilon^M=0.03$, and $0.0325$ have the same topology.
  }\label{fig:4beam-epsm}
\end{figure*}

In addition to the b-graph presented in Sec.~\ref{sec:fourunit} for 
$\varepsilon^M=0.035$, we map the response of sample B for $\varepsilon^M=0.03$, $0.0325$, $0.0375$, and $0.04$ (Fig.~\ref{fig:4beam-epsm}). In these responses, we find the same edge motifs and the same loop motifs as in Sec.~\ref{sec:fourunit}.
Note that the qualitative response for $\varepsilon^M=0.03$, and $\varepsilon^M=0.0325$ is equivalent (Fig.~\ref{fig:4beam-epsm}a-b).
Together, we conclude that high-dimensional b-graphs can be robustly mapped using our recursive strategy.

\section*{{APPENDIX F: GROOVY SHEET}} \label{sec:groovydetails}
{
Our groovy sheets are fabricated in a three-step process. First, the corrugations are formed by casting degassed Zhermack Double Elite 32 silicone rubber (Young's modulus of $\approx 1 \mathrm{MPa}$, Poisson's ratio of $\approx0.5$) into a two-part mold fabricated using UltiMaker S3 printers. 
After curing at room temperature, the sample is demolded and mounted in a clamp featuring matching corrugations, leaving both ends of the sheet exposed.
In the second and third step, a separate container is prepared with a thin layer of degassed silicone rubber. The exposed end of the sheet is then carefully lowered into this layer to create an end cap. 
After curing of the first end cap, this procedure is repeated for the opposite end. 
Together, these end caps stabilizes the 
corrugations and facilitates coupling to the experimental device.
}

{
The sample shown in Fig.~\ref{fig:GroovySheet} has a thickness $h=1.0\pm0.1\,\mathrm{mm}$, length $L=100\pm3\,\mathrm{mm}$, and features sinusoidal corrugations with with amplitude $A=8.5\pm0.1\,\mathrm{mm}$ and pitch $w=22.5\pm0.5\,\mathrm{mm}$.
The end caps have thickness $t=1.0\pm0.2\,\mathrm{mm}$.
}

{
To explore the strain space, we 
determined the irreversible transitions for four different paths:
$P_\mathrm{I}:(0,\!0)\!\rightarrow\!(\varepsilon^m,\!0)\!\rightarrow\!(\varepsilon^m,\varepsilon^M)\!\rightarrow\!
(0,\varepsilon^M)\!\rightarrow\!
(0,0)$; 
$P_\mathrm{II}:(0,\!0)\!\rightarrow\!(0,\varepsilon^m)\!\rightarrow\!(\varepsilon^M,\varepsilon^m)\!\rightarrow\!
(\varepsilon^M,0)\!\rightarrow\!
(0,0)$; 
$P_\mathrm{III}:(0,\!0)\!\rightarrow\!(\varepsilon^M,\!0)\!\rightarrow\!(\varepsilon^m,\!0)\!\rightarrow\!(\varepsilon^m,\varepsilon^M)$
$P_\mathrm{IV}:(0,\!0)\!\rightarrow\!(0,\varepsilon^M)\!\rightarrow\!(0,\varepsilon^m)\!\rightarrow\!(\varepsilon^M,\varepsilon^m)$,
where we
scan $\varepsilon^m$ between 0.002 and 0.05, and we fix $\varepsilon^M=0.054$.
Based on these explorations, we selected the three strain paths corresponding to the motifs shown in Fig.~\ref{fig:GroovySheet}. 
}

\end{document}